\documentclass[twocolumn]{aastex631}

\newcommand{\be}{\begin{equation}}
\newcommand{\ee}{\end{equation}}
\newcommand{\bea}{\begin{eqnarray}}
\newcommand{\eea}{\end{eqnarray}}
\newcommand{\bel}{\begin{align}}
\newcommand{\eel}{\end{align}}
\newcommand{\bsplit}{\begin{split}}
\newcommand{\esplit}{\end{split}}

\def\prf{q}

\def\params{\boldsymbol{\phi}}

\def\d{\text{d}}

\def\p{\partial}

\def\s{\boldsymbol{\epsilon}}
\def\es{\epsilon}
\def\sp{\epsilon_+}
\def\hsp{\hat{\epsilon}_+}
\def\spi{\epsilon_{+,i}}
\def\hspi{\hat{\epsilon}_{+,i}}

\def\hspt{\hat{\epsilon}_{+,t}}

\def\sm{\epsilon_-}

\def\smt{\epsilon_{-,t}}

\def\latent{z}
\def\normal{{\rm n}}
\def\f{\textbf{f}}
\def\CE{\mathcal{C}}

\usepackage{graphicx}	
\usepackage{amsmath}	
\usepackage{enumitem}
\usepackage{hyperref}

\begin{document}

\title{Short-Term Turbulence Prediction for Seeing Using Machine Learning}

\author{Mary Joe Medlej}
\affiliation{Université Côte d’Azur, Observatoire de la Côte d’Azur, J.L. Lagrange, Parc Valrose 06108 Nice
Cedex 2, France}
\author{Rahul Srinivasan}
\affiliation{SISSA, Via Bonomea 265, 34136 Trieste, Italy}
\affiliation{IFPU - Institute for Fundamental Physics of the Universe, Via Beirut 2, 34014 Trieste, Italy}
\author{Simon Prunet}
\affiliation{Université Côte d’Azur, Observatoire de la Côte d’Azur, J.L. Lagrange, Parc Valrose 06108 Nice
Cedex 2, France}
\author{Aziz Ziad}
\affiliation{Université Côte d’Azur, Observatoire de la Côte d’Azur, J.L. Lagrange, Parc Valrose 06108 Nice
Cedex 2, France}
\author{Christophe Giordano}
\affiliation{Université Côte d’Azur, Observatoire de la Côte d’Azur, J.L. Lagrange, Parc Valrose 06108 Nice
Cedex 2, France}

\email{mary-joe.medlej@oca.eu}

\begin{abstract}
Optical turbulence, driven by fluctuations of the atmospheric refractive index, poses a significant challenge to ground-based optical systems, as it distorts the propagation of light. This degradation affects both astronomical observations and free-space optical communications. While adaptive optics systems correct turbulence effects in real-time, their reactive nature limits their effectiveness under rapidly changing conditions, underscoring the need for predictive solutions. In this study, we address the problem of short-term turbulence forecasting by leveraging machine learning models to predict the atmospheric seeing parameter up to two hours in advance. We compare statistical and deep learning approaches, with a particular focus on probabilistic models that not only produce accurate forecasts but also quantify predictive uncertainty, crucial for robust decision-making in dynamic environments. Our evaluation includes Gaussian processes (GPs) for statistical modeling, recurrent neural networks (RNNs) and long short-term memory networks (LSTMs) as deterministic baselines, and our novel implementation of a normalizing flow for time series (\textsc{FloTS}) as a flexible probabilistic deep learning method. All models are trained exclusively on historical seeing data, allowing for a fair performance comparison. We show that \textsc{FloTS} achieves the best overall balance between predictive accuracy and well-calibrated uncertainty.
\end{abstract}

\section{Introduction}

Optical turbulence (OT) refers to the random fluctuations in the refractive index of the atmosphere caused by variations in temperature, pressure, and wind. These fluctuations distort the propagation of light through the atmosphere, posing a fundamental limitation to the performance of ground-based optical systems. OT affects many applications, from astronomy to free-space optical (FSO) communication, by degrading image quality or disrupting signal transmission.

In astronomical observations, particularly with extremely large telescopes (ELTs), turbulence introduces distortions in starlight as it passes through the Earth’s atmosphere, significantly reducing image quality, which is commonly quantified by the seeing parameter $\s$ \citep{roddier1981v,Tokovinin_2002}. The seeing parameter is defined as the full width at half-maximum of the long-exposure, seeing-limited point spread function (PSF) of a star image formed at the focus of a large-aperture telescope, primarily caused by atmospheric turbulence. Similarly, in FSO communication, turbulence causes signal scattering and intensity fluctuations, thereby reducing the reliability and efficiency of high-speed optical data transmission.

To compensate for the real-time effects of turbulence, adaptive optics (AO) systems have been widely deployed in both fields. AO dynamically corrects wavefront errors using deformable mirrors controlled by feedback from wavefront sensors \citep{roddier}. While AO has significantly improved system performance, it remains inherently reactive and limited by the unpredictability of turbulence over short timescales \citep{tyson,khaligi,wang}. This limitation has motivated growing interest in OT forecasting, which can enhance AO performance by enabling predictive corrections or proactive operational adjustments.

In astronomy, turbulence prediction is crucial for optimizing observation scheduling and improving AO performance, thereby maximizing the scientific output of telescopes. Long-term predictions support observatory planning, including telescope scheduling and resource allocation, by providing insight into upcoming observing conditions. Accurate short-term predictions, on the other hand, enable real-time adjustments to AO systems and observation strategies, ensuring optimal image quality under changing atmospheric conditions.
In FSO communication, turbulence prediction is equally important. Long-term predictions assist with network planning and resource optimization, ensuring reliable connectivity over extended periods. Short-term forecasts allow real-time adaptation of system parameters, such as power control and beam tracking, to maintain stable communication links \citep{Khalighi,Majumdar2015}. Together, long-term and short-term predictions play complementary roles in addressing the challenges posed by atmospheric turbulence.

Traditional numerical simulations based on physical models of atmospheric dynamics have produced promising results for long-term turbulence prediction \citep{Frehlich, Christophe_2013, Cherubini_2013, Christophe_2014, masciadri2017optical, Rafalimana_2019, Lyman, Rafalimana_2020, Christophe_2021, Rafalimanana_2022}. However, their high computational cost makes them unsuitable for rapid short-term forecasting, highlighting the need for faster predictive approaches.

Machine Learning (ML) methods offer strong predictive capabilities combined with fast inference, making them particularly well-suited for modeling the complex and rapidly evolving nature of atmospheric turbulence. In this work, we focus on short-term turbulence prediction and investigate ML models for forecasting the seeing parameter with prediction horizons of up to two hours. By leveraging modern data-driven approaches, we aim to improve the accuracy and efficiency of short-term turbulence forecasting for both astronomical observations and FSO communication systems.

Previous studies have demonstrated the potential of ML approaches for atmospheric seeing prediction. \citet{2022data} investigated several data-driven techniques for this task. \citet{Cherubini_2021} applied ML methods to forecast seeing conditions for the following five nights. \citet{Giordano_ML} used the Random Forest (RF) algorithm to predict seeing with a lead time of up to two hours, while \citet{Masciadri_1,Masciadri_2} combined autoregressive models and RF for seeing prediction. Similarly, \citet{Milli} explored multilayer perceptrons and RF for short-term forecasts within a two-hour window. \citet{Kornilov_2015} treated seeing prediction as an autoregressive problem and applied a linear autoregressive integrated moving average (ARIMA) model.

Building upon these works, our study compares two main categories of models for short-term seeing prediction: statistical models and deep learning models. A key aspect of this work is the inclusion of probabilistic models within both categories. Unlike deterministic approaches that provide only a single forecast value, probabilistic models also estimate the uncertainty associated with predictions, which is particularly important in atmospheric sciences due to the intrinsic variability of the atmosphere. To evaluate the benefits of probabilistic forecasting, deterministic models are also included for comparison.

Within the statistical category, we employ Gaussian Processes (GP), a probabilistic model that naturally provides both predictions and confidence intervals (Section~\ref{sec:gp}). In the deep learning category, we consider both deterministic and probabilistic architectures. Recurrent neural networks (RNNs) and long short-term memory networks (LSTMs) serve as deterministic baselines designed to capture temporal dependencies in sequential data (Sections~\ref{sec:rnn} and \ref{sec:lstm}). For probabilistic deep learning, we use a normalizing-flow-based model for time series, hereafter referred to as \textsc{FloTS}, which is capable of modeling complex non-Gaussian distributions (Section~\ref{sec:flow}).

A normalizing flow is a class of generative neural networks that models a probability distribution by transforming a simple base distribution (typically a standard Gaussian) into a more complex target distribution through a sequence of invertible and differentiable mappings. This invertible transformation also enables sampling from the learned distribution by drawing samples from the latent space and mapping them back to the data space.

By incorporating GP and \textsc{FloTS}, our framework integrates probabilistic modeling into both statistical and deep learning approaches. Within the autoregressive setting considered in this work, these models provide point forecasts together with estimates of predictive uncertainty, which can support more informed decision-making in atmospheric applications.

To ensure a fair comparison, all models are trained using only historical seeing measurements without incorporating additional atmospheric or instrumental variables. This minimalist setup isolates the predictive capability of each model under identical data conditions. Including additional features such as temperature, wind, or humidity would increase both complexity and stochasticity, making information integration more challenging \citep{2022data}. We leave the exploration of such multivariate models to future work.

The remainder of this paper is organized as follows. Section~\ref{sec:data} introduces the dataset and preprocessing steps used for model training. Section~\ref{sec:methods} describes the forecasting models and their architectures, including the selection of optimal input lengths (Section~\ref{sec:training_length}) and the calibration procedure (Section~\ref{sec:calibration}). Section~\ref{sec:results} presents the statistical evaluation of the models and provides case studies illustrating model behavior in specific forecasting scenarios. Finally, Section~\ref{sec:conclusion} summarizes the main findings of this study.

\section{Data Acquisition and Preparation}
\label{sec:data}
\subsection{Data Acquisition}
\label{sec:data_acqui}
The Maunakea Weather Center (MKWC) provided the data used in this study. Observations were collected using a Monitoring Device (MD), which combines the Multi-Aperture Scintillation Sensor (MASS) and the Differential Image Motion Monitor (DIMM) \citep{vernin1995measuring,kornilov}. The MD is located at the summit of Maunakea, a site globally recognized for its exceptional astronomical observing conditions, including dark skies due to minimal light pollution, excellent seeing, low humidity, high elevation (4,205 meters), and a position above most atmospheric water vapor. These characteristics make Maunakea one of the most important astronomical sites in the world.
The MD is installed at the top of the Canada-France-Hawaii Telescope (CFHT) instrument tower, approximately 7 meters above the ground \citep{Lyman}. It is strategically positioned between the Gemini and CFHT observatories. It is minimally affected by terrain or nearby structures when winds originate from the west or east \citep{Silva}. 
The data have been available since mid-2009 and are continuously updated. Real-time MD data are publicly available through the Maunakea Weather Center website\footnote{\url{http://mkwc.ifa.hawaii.edu/current/seeing/}}.
In this study, we used DIMM seeing data between September 2009 and November 2024. DIMM measures the integrated optical turbulence across the entire atmosphere, estimating the total atmospheric seeing. The seeing data range on the Maunakea summit varies between 0.04$^{\prime\prime}$ and 5$^{\prime\prime}$.

\subsection{Data Preparation}
Most standard ML models, including RNNs and LSTMs, assume that the input sequence is sampled at regular time intervals \citep{lipton2015learning,che2018recurrent}. This assumption is inherent to the recursive formulation of these models, where each time step corresponds to a uniform update in time \citep{Goodfellow-et-al-2016}. In our case, we chose a sampling rate of 10 minutes, which aligns with the practical needs of astronomical operations and FSO communication systems \citep{Turchi_2022,Masciadri_2}. High-frequency variability on shorter timescales is generally less relevant for seeing prediction, since such rapid fluctuations fall outside the response timescales of observatory operations and the real-time adaptability of FSO systems. Astronomers prioritize understanding trends, whether the seeing is increasing, decreasing, or remaining stable, so they can make informed decisions about adjusting observation modalities or scientific programs. In FSO communication, this interval similarly captures turbulence trends critical for maintaining signal quality, enabling effective real-time adjustments.

The choice of a 10-minute grid therefore represents a compromise between preserving short-term seeing evolution and reducing the impact of very high-frequency fluctuations that are less relevant for the forecasting horizon considered here. We used local linear regression rather than a simple bin average or median because the original measurements are not always exactly aligned with the desired grid points. A local linear fit uses the timing of the measurements within the bin and estimates the seeing value at the target time, rather than assigning the same aggregate value to the entire interval. This is useful when the seeing evolves monotonically or approximately linearly over a short window, since it preserves local trends while regularizing the sampling. A simple bin-wise median would be less sensitive to intermittent spikes, but it would not explicitly use the sub-bin temporal ordering and could flatten short-term increases or decreases that are relevant for autoregressive forecasting. More sophisticated, robust alternatives, such as robust local linear regression or Huber-type regression, could, in principle, reduce the influence of outliers while still preserving the relative timing of the measurements within each bin.

To prepare the atmospheric seeing data for machine learning modeling, we performed temporal interpolation and resampling to regularize the time series before splitting it into training, validation, and testing sets. The raw seeing measurements were recorded at 90-second intervals but occasionally contained gaps, which pose challenges for models that require fixed time steps. To address this, we interpolated the data onto a uniform time grid with a fixed 10-minute sampling interval.

We note, however, that the resampling strategy can influence the effective smoothing of the time series, especially because seeing can exhibit intermittent spikes and non-Gaussian variability. For this reason, we avoid interpolating across large gaps by first segmenting the data into continuous blocks, and we restrict the interpolation to local windows around each target time. Although a full sensitivity study comparing local linear interpolation with simple bin-wise aggregation and robust time-aware alternatives is left for future work, we expect the main qualitative conclusions of the present study to be robust to reasonable resampling choices, since the models are evaluated on two-hour trends rather than individual 90-second fluctuations.

The resampling and interpolation procedure consisted of the following steps:
\begin{enumerate}[label=\roman*]
\item Segmentation of continuous time blocks: We calculate the time difference between consecutive pairs of measurements. A new segment is initiated when the gap between measurements exceeds 10 minutes. Hence, every segment contains a continuous block of data without large gaps.

\item Definition of target grid: Within each segment, we create a time grid with a 10-minute interval. These time steps are the targets for interpolation.

\item Local bin interpolation with linear regression: Around each target time step, we defined a bin centered at that point, extending one sampling interval on either side. Data points within each bin are used to estimate the interpolated value at the center. We performed linear regression on the bin's time indices, converted to seconds relative to the bin start, and their corresponding seeing values. The regression model was then evaluated at the target time to obtain the interpolated value. This procedure preserves the local temporal ordering of the measurements inside the bin, while avoiding interpolation across discontinuous observing periods.

\item Assembly of the interpolated dataset: All interpolated values were compiled into a new time series with uniform 10-minute spacing. This regularized dataset ensures compatibility with machine learning models that require fixed-size input windows and consistent temporal structure.
\end{enumerate}
For the LSTM, RNN, and \textsc{FloTS} models, each continuous segment of data was processed independently. Given sufficiently large continuous lengths, the data are segmented into input-output pairs, denoted by $\sm \in \mathbb{R}^{d_-}$ and $\sp \in \mathbb{R}^{d_+}$, corresponding to sequence lengths $d_-$ and $d_+$, respectively. The input $\sm$ represents a sequence of historical seeing measurements used for prediction, while $\sp$ is the corresponding future window that the model aims to forecast. The slicing was performed using a sliding window mechanism with a single timestep stride, ensuring that all valid subsequences within a group were extracted. We evaluated different input sequence lengths, testing durations from 6 hours (corresponding to 36 time steps) down to 2 hours (12 time steps), as described in Section \ref{sec:training_length}. The output horizon $\sp$ was fixed at 2 hours (12 time steps), aligned with the predictive goal of the study.

After generating the complete set of $(\sm, \sp)$ pairs, a total of 71\,906 pairs were obtained. 
Each pair corresponds to a two-hour sequence composed of 12 time steps with a temporal resolution 
of 10 minutes. The dataset therefore represents approximately 143\,812 hours of time series data 
(about 5\,992 days).

The resulting pairs were randomly shuffled and split into 81.5\% for training (58\,603 pairs), 
8.5\% for validation (6\,112 pairs), and 10\% for testing (7\,191 pairs). This randomization 
was applied across all valid sequences, ensuring statistical consistency and reducing potential 
bias in performance evaluation. For the GP model, which operates on fixed input-output pairs without requiring sequence history or internal state memory, the same data preprocessing steps were followed to ensure comparability. Each ($\sm$, $\sp$) pair was constructed according to the same input and output lengths used for the deep learning models. However, unlike the deep learning models, the GP implementation does not require a fixed
training/validation/test split to operate. Instead, random batches were sampled
from the $\sm$ dataset to estimate the GP hyperparameters. These hyperparameters were
then kept fixed and used to generate predictions for each batch individually, producing
the corresponding $\sp$ forecasts (Section~\ref{sec:gp}).
In principle, separating the data used for hyperparameter learning from the data used
for prediction also helps prevent overfitting in GP models, as in other machine learning
approaches.

This approach preserves the temporal structure of the original data while enforcing consistent time steps. The local linear regression within sliding bins provides a simple and robust way to handle nonuniform sampling while avoiding interpolation across large observational gaps.
It is also noteworthy that no normalization was applied to the input data. Since the only feature used in the model is the seeing value, which naturally ranges between approximately 0.04$^{\prime\prime}$ and 5$^{\prime\prime}$ (Section~\ref{sec:data_acqui}), the scale of the input is already constrained and physically meaningful. Normalization is typically applied when features have different units, scales, or orders of magnitude to ensure numerical stability during model training. In this case, the seeing values are homogeneous, bounded, and consistent across samples, allowing the models to learn effectively without additional rescaling. 

We also monitored the training behavior and did not observe numerical instabilities, exploding losses, or convergence difficulties attributable to the lack of normalization. The use of a single physically meaningful variable further reduces the risk of scale imbalance between inputs. We therefore kept the seeing values in their original units, which also facilitates the interpretation of the model predictions and uncertainty estimates directly in arcseconds. Nevertheless, normalization or standardization could be considered in future extensions involving additional meteorological predictors with different physical units and dynamical ranges.

\section{Methods}
\label{sec:methods}
The goal of this study is to predict atmospheric seeing using forecasting techniques. Atmospheric seeing is affected by turbulent fluctuations in the atmosphere, which are both random and time-dependent. Accurately forecasting seeing therefore requires models that can process sequential data, capture complex temporal dynamics, and, ideally, provide uncertainty estimates to quantify prediction confidence.

We first investigated recurrent neural network architectures, namely RNNs and LSTMs \citep{Alex}, which are widely used for sequential forecasting tasks. RNNs process input data step by step, using prior outputs to inform future predictions \citep{RNN}. LSTMs extend this capability by retaining information over longer time horizons \citep{lstm_1997,LSTM_book}, making them particularly effective for modeling temporal dependencies in atmospheric turbulence. These models showed good predictive performance when trained on past seeing measurements. However, both approaches are deterministic and produce only point predictions, without providing a measure of uncertainty. This limitation can be problematic in operational contexts such as telescope scheduling or optical communication, where reliable confidence intervals are important.

To address this limitation, we also considered Gaussian Processes (GPs), which are inherently probabilistic models. GPs provide both a prediction and an associated confidence interval \citep{GP_book}, making them well-suited to scientific applications where uncertainty estimation is essential. They are particularly effective for modeling smooth trends and perform well on relatively small datasets. However, GPs also have limitations: they assume Gaussian-distributed data (which may not always hold for atmospheric turbulence) \citep{jones2004non}, and their computational cost scales poorly with increasing dataset size.

To overcome these limitations, we incorporate normalizing flows in our analysis. The \textsc{FloTS} model is a generative and probabilistic deep learning approach that transforms a simple base distribution (e.g., Gaussian) into a complex target distribution through a sequence of invertible mappings \citep{flow_1}. This allows \textsc{FloTS} to capture non-Gaussian characteristics in the data while providing full probabilistic predictions. In addition, \textsc{FloTS} scales more efficiently with dataset size than GPs, making it a promising candidate for high-resolution or operational turbulence forecasting.

In summary, this study compares three types of models for atmospheric seeing prediction:

\begin{itemize}
    \item Deterministic deep learning models: RNN and LSTM.
    \item Probabilistic statistical model: Gaussian Processes (GP).
    \item Probabilistic deep learning model: Normalizing Flow (\textsc{FloTS}).
\end{itemize}
We now describe the architecture and implementation details of each model we used in this analysis.

\subsection{Recurrent Neural Network (RNN)}
\label{sec:rnn}
RNNs are a class of neural networks specifically designed to process sequential data by capturing temporal dependencies. Unlike traditional feedforward networks such as multilayer perceptrons, RNNs incorporate hidden states that enable information to persist across time steps. This is accomplished through a recursive structure, where the hidden state vector at time step \(t\), denoted by \(h_t\), is computed as

\begin{equation}
h_t = \sigma\left( W_h h_{t-1} + W_{\sm} \smt + b_h \right),
\end{equation}

where \( \smt \) is the input at time step \(t\), \(h_{t-1}\) is the previous hidden state vector, \(W_h\) and \(W_{\sm}\) are weight matrices, \(b_h\) is a bias term, and \(\sigma\) denotes a nonlinear activation function. In the encoder–decoder framework used in this work, \(h_t\) denotes the hidden state of the encoder RNN, which processes the input sequence and summarizes it into a latent representation. 

The decoder then generates the forecast sequence using its own hidden state \(g_t\), which is conditioned on the final encoder state \(h_T\). The decoder state evolves as
\begin{equation}
g_t = \sigma\left( W_g g_{t-1} + W_c h_T + W_y \hat{\epsilon}_{t-1} + b_g \right),
\end{equation}

where $W_g$, $W_c$, and $W_y$ are weight matrices and $b_g$ is a bias term. 
The term $\hat{\epsilon}_{t-1}$ denotes the predicted seeing value at the previous 
time step. This autoregressive dependency allows the decoder to propagate 
temporal information through the generated forecast sequence. The predicted seeing value at time step \(t\) is obtained by applying a linear projection to the decoder hidden state:

\begin{equation}
\hspt = W_{\sp} g_t + b_{\sp}.
\end{equation}

In this study, RNNs were employed as an initial approach for modeling the temporal dynamics of atmospheric turbulence. The architecture consists of an encoder RNN layer with 100 hidden units, followed by two decoder RNN layers, each also comprising 100 hidden units, which further refine and propagate the encoded temporal information \citep{cho2014,pascanu2014}. To mitigate overfitting, a dropout layer with a rate of 0.3 is inserted between the encoder and decoder layers.
The model is trained using the Mean Squared Error (MSE) loss function, defined as:
\begin{equation}
\mathcal{L}_{\text{MSE}} = \frac{1}{N} \sum_{i=1}^{N} \left( \spi - \hspi \right)^2
\end{equation}
where \( \spi \) and \( \hspi \) represent the true and predicted seeing values, respectively. Optimization is performed using the Adam algorithm with an initial learning rate of \( 10^{-4} \). A learning rate scheduler is employed to reduce the rate by a factor of 0.5 if the validation loss plateaus for five consecutive epochs.
Although effective at capturing short-term dependencies, RNNs suffer from the vanishing gradient problem \citep{hochreiter1998vanishing}. During backpropagation through time (BPTT), the loss can be written as a sum over time steps, \(\mathcal{L} = \sum_k \mathcal{L}_k\). The gradient of a future loss term with respect to a hidden state at an earlier time step \(t\) (with \(k > t\)) follows from the chain rule:

\begin{equation}
\frac{\partial \mathcal{L}_k}{\partial h_t} =
\frac{\partial \mathcal{L}_k}{\partial h_k}
\frac{\partial h_k}{\partial h_t}.
\end{equation}

Because the hidden states are computed recursively, the term 
\(\frac{\partial h_k}{\partial h_t}\) involves repeated multiplications of the recurrent 
weight matrix \(W_h\). As a result, the gradient contains products of the form

\begin{equation}
\frac{\partial h_k}{\partial h_t}
\propto
\prod_{i=t+1}^{k} W_h.
\end{equation}

If the spectral radius (largest eigenvalue) of \(W_h\) is smaller than one, these repeated multiplications cause the gradient to decay exponentially as \(k-t\) increases, making it difficult for the network to learn long-term dependencies.This limitation is particularly detrimental when modeling sequences that require memory of distant past events. To address this issue, more advanced architectures such as Long Short-Term Memory (LSTM) networks introduce gating mechanisms that regulate the flow of information and mitigate gradient decay \citep{hochreiter1997long}. Accordingly, this study transitions to LSTM-based models to assess whether they provide tangible benefits over standard RNNs for seeing prediction, especially in the presence of long-range temporal dependencies.

\begin{table}
\centering
\includegraphics[width=80mm,scale=1]{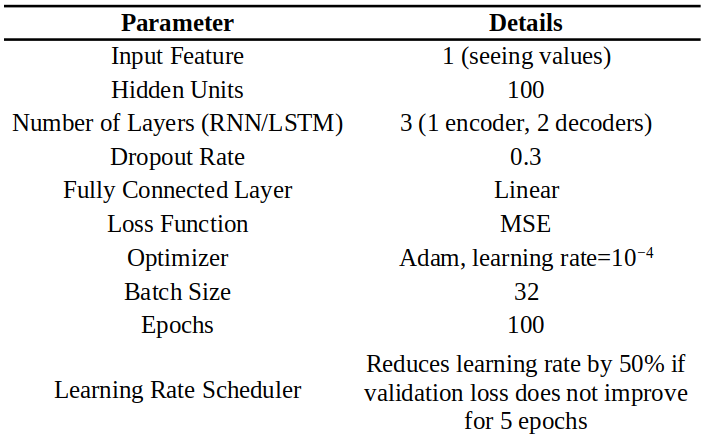}
\caption{Model Architectures and Hyperparameters for RNN and LSTM.}
\label{table:parameters}
\end{table}

\subsection{Long Short-Term Memory (LSTM)}
\label{sec:lstm}

The LSTM model implemented in this study is designed to predict atmospheric seeing by capturing potential long-term dependencies in sequential turbulence data. LSTMs extend standard RNNs by addressing the vanishing gradient problem, which limits the ability of conventional RNNs to retain information over long sequences. By incorporating an internal gating mechanism, LSTMs regulate the flow of information, enabling the selective storage of relevant past data while discarding less useful details. This property makes them particularly suitable for time series forecasting, where dependencies between past and future observations may play a crucial role.

Unlike standard RNNs, which rely solely on recurrent connections, LSTMs introduce three gates (Equation~\ref{eq:LSTM_gates}) that govern information flow. The following equations describe the operations of a single LSTM cell, which is used in both the encoder and decoder layers of the architecture. The forget gate determines the extent to which past information should be retained or discarded. The input gate controls how much new information is added to the internal memory, while the output gate regulates how much of the stored information contributes to the current hidden state. These mechanisms allow the model to dynamically adjust the influence of past observations, mitigating the issues associated with long-term dependencies in sequential data:

\begin{equation}
\begin{aligned}
f_t &= \sigma\left( W_f \left[ h_{t-1}, \smt \right] + b_f \right),\\
i_t &= \sigma\left( W_i \left[ h_{t-1}, \smt \right] + b_i \right),\\
o_t &= \sigma\left( W_o \left[ h_{t-1}, \smt \right] + b_o \right)
\end{aligned}
\label{eq:LSTM_gates}
\end{equation}

where \( f_t, i_t, o_t \) represent the forget, input, and output gates, respectively; \( h_{t-1} \) is the hidden state from the previous time step; \( \smt \) is the input at time \( t \); and \( W_f, W_i, W_o \) and \( b_f, b_i, b_o \) are learnable weight matrices and bias terms. The sigmoid activation function \( \sigma \in (0,1) \) allows each gate to control the extent of information flow.

The cell state \( c_t \), which stores long-term information, is updated using

\begin{equation}
\tilde{c}_t = \tanh\left( W_c \left[ h_{t-1}, \smt \right] + b_c \right)
\label{eq:cell_candidate}
\end{equation}

\begin{equation}
c_t = f_t \odot c_{t-1} + i_t \odot \tilde{c}_t
\label{eq:cell_update}
\end{equation}

where \( \tilde{c}_t \) is the candidate cell state, \( c_{t-1} \) is the previous cell state, and \( \odot \) denotes element-wise multiplication.

The hidden state is then computed by modulating the updated cell state
through the output gate:

\begin{equation}
h_t = o_t \odot \tanh(c_t)
\label{eq:hidden_update1}
\end{equation}

where \( o_t \) is the output gate. This operation controls how much of
the internal memory stored in the cell state contributes to the hidden
representation passed to the next time step or subsequent network layers.

In the encoder–decoder architecture used in this study, the decoder is conditioned on the final encoder hidden state \( h_T \), which summarizes the input sequence. 
The decoder LSTM follows the same gating structure as the encoder.
However, instead of receiving the past seeing measurements
\( \epsilon_{-,t} \) as input, the decoder operates autoregressively
and takes as input the previously generated prediction
\( \hat{\epsilon}_{+,t-1} \). Consequently, the decoder updates its
internal states using the same LSTM equations, with the input
\( \hat{\epsilon}_{+,t-1} \) replacing \( \epsilon_{-,t} \).

The decoder hidden state \( g_t \) is obtained from the decoder cell
state \( d_t \) through

\begin{equation}
g_t = o_t \odot \tanh(d_t),
\label{eq:hidden_update2}
\end{equation}

where \( o_t \) is the output gate of the decoder LSTM.

The predicted seeing value is obtained by applying a linear projection to the decoder hidden state:

\begin{equation}
\hspt = W_{\sp} g_t + b_{\sp}
\label{eq:lstm_prediction}
\end{equation}

While LSTMs are designed to capture long-term dependencies, it remains uncertain whether such dependencies are present in atmospheric seeing data. Therefore, their effectiveness relative to standard RNNs remains to be evaluated for this dataset.

The architecture of the LSTM model follows the same structure as the RNN model described previously, with LSTM layers replacing the standard RNN layers. The hyperparameters and optimization strategies, including batch size, learning rate, dropout rate, and loss function, are consistent with those used for the RNN model and are summarized in Table~\ref{table:parameters}.

\subsection{Gaussian Process (GP)}
\label{sec:gp}
GPs are employed in this study to predict atmospheric seeing, providing uncertainty estimates for the predictions. Unlike deterministic models, GPs are non-parametric probabilistic models that define a distribution over functions \citep{GP_book,seeger2004gaussian}. Specifically, a GP assumes that any finite collection of inputs, the corresponding function values are jointly distributed according to a multivariate Gaussian distribution, enabling the model to produce both a predictive mean and an associated confidence interval at each new input.
Formally, a GP is defined by a mean function \( \mu(t) \) and a covariance function (kernel) \( k(t, t') \), such that:
\begin{equation}
  f(t) \sim \mathcal{GP}\big(\mu(t), \, k_\theta(t,t')\big)
  \label{eq:latent}
\end{equation}
where $\theta$ collects the kernel hyperparameters (length scale, signal variance, noise).\\

Let \( \mathcal{D} = \{ \sm^{i}, \sp^{i} \}_{i=1}^{N} \) denote the training dataset, where \( N \) is the total number of training samples, \( \sm^{i} \in \mathbb{R}^{d_-} \) is the input seeing vector (for $d_-$ time steps) and \( \sp^{i} \in \mathbb{R}^{d_+} \) is the corresponding output values (for $d_+$ time steps). Under the GP prior, the key assumption is that the joint distribution of the input and output vectors can be modeled as a multivariate Gaussian distribution, with a covariance structure governed by a kernel function that encodes similarity between input points.

\begin{equation}
\begin{bmatrix}
\sm \\
\sp
\end{bmatrix}
\sim \mathcal{N} \left(
\begin{bmatrix}
\boldsymbol{\mu}_- \\
\boldsymbol{\mu}_+
\end{bmatrix},
\begin{bmatrix}
K_{--} + \sigma_n^2 \mathbf{I} & K_{-+}^\top \\
K_{-+} & K_{++}
\end{bmatrix}
\right)
\end{equation}

where, $\mu_-$ and $\mu_+$ are the mean functions for the input and output values and 

\begin{equation}
\begin{aligned}
K_{--} &= k(t_-, t_-) \in \mathbb{R}^{d_- \times d_-} \\
K_{-+} &= k(t_+, t_-) \in \mathbb{R}^{d_+ \times d_-} \\
K_{++} &= k(t_+, t_+) \in \mathbb{R}^{d_+ \times d_+}
\end{aligned}
\end{equation}

Here, $k(t, t^{\prime})$ denotes the kernel function, which determines the correlation between time steps $t$ and $t^\prime$. As before, $t_-$ and $t_+$ denote time steps of the input and output, respectively. $\mathbf{I}$ is the \( d_- \times d_- \) identity matrix and \( \sigma_n^2 \) is the variance of the Gaussian observation noise, which accounts for measurement errors or stochastic fluctuations in the data.

The output seeing $\sp$ given the input values $\sm$ therefore follows a conditional Gaussian distribution given by,
\begin{equation}
\sp \mid \sm \sim \mathcal{N}(\boldsymbol{\mu}_{\sp|\sm}, \Sigma_{\sp|\sm}),
\end{equation}
with 
\begin{equation}
\begin{aligned}
\boldsymbol{\mu}_{\sp|\sm} &= \boldsymbol{\mu}_+ + K_{-+} \left(K_{--} + \sigma_\epsilon^2 \mathbf{I}\right)^{-1} (\sm - \boldsymbol{\mu}_-),\\
\Sigma_{\sp|\sm} &= K_{++} - K_{-+} \left(K_{--} + \sigma_\epsilon^2 \mathbf{I}\right)^{-1} K_{-+}^\top \,.
\end{aligned}
\end{equation}
This formulation allows GPs to output both the predicted seeing value and an associated standard deviation that quantifies prediction uncertainty. In our study, we use the exponential kernel:
\begin{equation}
  k_\theta(t,t') = \sigma^2 \exp\!\left(-\frac{|t - t'|}{l}\right),
\end{equation}
where \( \sigma^2 \) is the variance, \( l \) is the length-scale. This kernel was selected for its ability to model the decaying temporal correlation present in turbulence data. When \( t \) and \( t^\prime \) correspond to nearby time steps, the correlation between \( \s \) and \( \s^\prime\) is high, and as the time difference increases, the correlation decays exponentially. To make GP training computationally manageable, a random subset of 500 $(\sm,\sp)$ pairs was selected from the training set. During hyperparameter optimization, only the input sequences $\sm$ are used to fit the GP model, while the corresponding targets $\sp$ are used later for evaluating the predictions. This choice balances computational cost and model accuracy, since computing the posterior involves inverting an \( d_- \times d_- \) matrix at a cost of \( \mathcal{O}(d_-^3) \). Using these batches, the GP learned its hyperparameters: length scale, variance, and noise variance. The length scale \( l \) controls the degree of temporal correlation between inputs, the signal variance \( \sigma^2 \) represents the amplitude of function variation, and the noise variance \( \sigma_n^2 \) accounts for measurement error. Hyperparameters learned across batches were averaged to ensure robust and stable behavior.

In our GP formulation (Equation~\ref{eq:latent}), the latent process $f(t)$ is defined
by a mean function $\mu(t)$ and a covariance function $k_\theta(t,t')$, where the
mean captures slow baseline variations and the kernel models correlated
fluctuations around that baseline. This distinction is important because the GP
is trained on many short seeing windows sampled throughout the year, which may
exhibit different baseline seeing levels corresponding to different atmospheric
regimes. Imposing a single fixed mean across all windows would therefore force
the kernel to absorb these offsets and could bias the inferred hyperparameters.
To avoid this, we apply per-window demeaning: for each window we subtract its
empirical mean and fit a zero-mean GP to the residuals, adding back the window
baseline at prediction time.

We also tested mean marginalisation by integrating out a constant offset under a
flat prior and obtained nearly identical accuracy and uncertainty. This is
expected in our setting because, over a few-hour window, the mean is well
approximated by a constant and forecasting is dominated by the kernel’s
short-term correlation structure.
While GPs provide a principled probabilistic framework for forecasting, their predictive distribution is Gaussian and therefore fully characterized by a mean function and a covariance matrix. This assumption may limit the ability of the model to represent asymmetric, heavy-tailed, or multi-modal predictive distributions. In addition, we emphasize that the GP results reported in this work reflect the specific GP configuration adopted here, rather than an inherent limitation of Gaussian Process methods in general. In particular, our implementation relies on a simple stationary exponential kernel with a single characteristic time scale, and the hyperparameters are estimated from a restricted subset of 500 training pairs for computational efficiency. These choices may limit the ability of the GP baseline to represent the multi-scale and potentially non-stationary structure of atmospheric seeing.

More expressive GP formulations could be considered in future work, including sums or products of kernels, multi-scale kernels, spectral mixture kernels, non-stationary kernels, or sparse/inducing-point approximations that allow GP models to scale to larger training sets. Such extensions could improve the flexibility of the GP framework while retaining its interpretability and probabilistic foundation. In the present study, the GP should therefore be interpreted as a simple and interpretable probabilistic baseline. We therefore also investigate normalizing flows, a class of probabilistic deep learning models capable of representing more flexible, non-Gaussian predictive distributions.

\subsection{Normalizing Flows (\textsc{FloTS})}
\label{sec:flow}
Given samples from the target distribution $\s \sim p(\s)$,
the normalizing flow learns a bijection map $\f$ such that $\s =\f_{\params}(z)$, where $z$ are samples from the latent Gaussian distribution and $\params$ are the network weights to be optimized. Then, the target distribution can be mapped by a change of basis of the latent distribution, 
\be
\label{eq:flow}
p(\s) \mapsto \prf_{\params}(\s) =  \normal(\f_{\params}^{-1}(\s))\, \left|{\rm det}\frac{\p\f_{\params}^{-1}}{\p\s}(\s) \right|\,,
\ee
where $\f_{\params}^{-1}$ is the inverse of $\f_{\params}$ ($z =\f_{\params}^{-1}(\s)$)
and ${\p \f_{\params}^{-1}}/{\p\s}$ is its corresponding Jacobian. A neural network models the invertible transformation $\f$ which can be made arbitrarily expressive, limited only by the expressivity of the network. In theory, any target distribution $p(\s)$ can be generated from the base distribution $\normal(z)$ under reasonable assumptions~\citep{Villani:2003, Bogachev:2005, Medvedev:2008}. 
 
The normalizing flow parameters $\params$ are optimized using the cross-entropy loss defined as, 
\be
\label{eq:standard-loss} 
\begin{split}
L(\params) &= -\int_{\mathbb{R}^d} p(\s) \, \log(\prf_{\params}(\s))\,\d\s\,\\
&=-{\rm E}_{p(\s)}\left[ \log(\prf_{\params}(\s))\right],
\end{split}
\ee
where ${\rm E}_{p(\s)}$ represents the expectation value for samples of $\s\sim p(\s)$. Thus, given a set of independent samples from the target distribution, the flow maximizes its predicted log-probability of the samples. The flow can be extended to learn a conditional probability distribution by passing the conditional information, often referred to as \textit{context}, as input to the neural network. 

Given observations of the seeing for past and future timesteps, $\{\sm,\,\sp\}$, we train a normalizing flow to model the probability distribution of future seeing values conditioned on past observations, $p(\sp|\sm)$. The trained flow can then be used to produce predictions of future seeing values sampled from the conditional distribution i.e., $\hsp \sim p(\sp|\sm)$.

In this implementation, we use a class of normalizing flows called masked autoregressive flow (MAF, \cite{MAF}). The MAF implementation of NF expands $p(\s)$ as the product of autoregressive conditional probabilities:
\be
p(\s_{d:1})=p(\s_d|\s_{d-1:1})\times p(\s_{d-1}|\s_{d-2:1}) \times ... \times p(\s_1),
\ee

where $\s_{d:1}$ is the $d$-dimensional vector (first $d$ time-steps). Conditioning on the past values $\sm$, $p(\es_{+,\,d:1}|\sm)$ is computed in MAF as:

\be
\begin{split}
p(\es_{+,\,d:1}|\sm) =& \,p(\es_{+,\,d}|\es_{+,\,d-1:1},\sm) \\
&\times p(\es_{+,\,d-1}|\es_{+,\,d-2:1},\sm) \\
& \times ... \times p(\es_{+,\,1}|\sm).
\end{split}
\ee

To this end, $\sm$ is passed to the MAF through a context encoder $\CE(\sm)\in\mathbb{R}^{2d_+}$ to identify the $d_+$-dimensional mean and standard deviation of the normal distribution in the latent space, i.e., $\latent\sim \normal_{\CE(\sm)}(\latent)$, where $\normal_{\CE(\sm)}:=\mathcal{N}(\CE_{d:1}(\sm),\CE_{2d:d+1}(\sm))$. The flow models the transformation $\f_{\params}$ that satisfies $\sp =\f_{\params}(\latent)$, and approximates $p(\sp|\sm)$ by $\prf_{\params}(\sp|\sm)$:

\be
\begin{split}
p(\sp|\sm) \mapsto & \, \prf_{\params}(\sp|\sm) \\
& =  \normal_{\CE(\sm)}(\f_{\params}^{-1}(\sp))\, \left|{\rm det}\frac{\p\f_{\params}^{-1}}{\p\sp}(\sp) \right|\,.
\end{split}
\ee

The encoder $\CE$ can be any tunable function of $\sm$ whose parameters are trained along with the MAF by the optimizer. We find the best results with an LSTM encoder, outperforming other variants we explored, including a simple linear network, a fully connected network (FCN), an RNN, and a GP.

\subsection{Hyperparameter tuning}
\label{sec:training_length}
An essential step in training our ML algorithms involved determining the optimal hyperparameters that describe the model. In the case of neural network architectures, the hyperparameters include the number and shapes of the layers of the network. For GPs, these correspond to the choice of kernels and their parameters. Another important common hyperparameter is the model-specific training length to achieve the best predictions for the 2-hour forecast. We select the optimal hyperparameters based on empirical performance. To assess the accuracy of these predictions, we evaluated various error metrics, including Root Mean Square Error (RMSE) and the Pearson correlation coefficient (\( r \)). In these equations, \( N \) represents the sample size, while \( X_i \) and \( Y_i \) denote the predicted values from the ML models and the corresponding measurements, respectively. \( \bar{X} \) and \( \bar{Y} \) are the means of the variables \( X \) and \( Y \). 

\begin{figure*}
\centering
\includegraphics[width=180mm,scale=1]{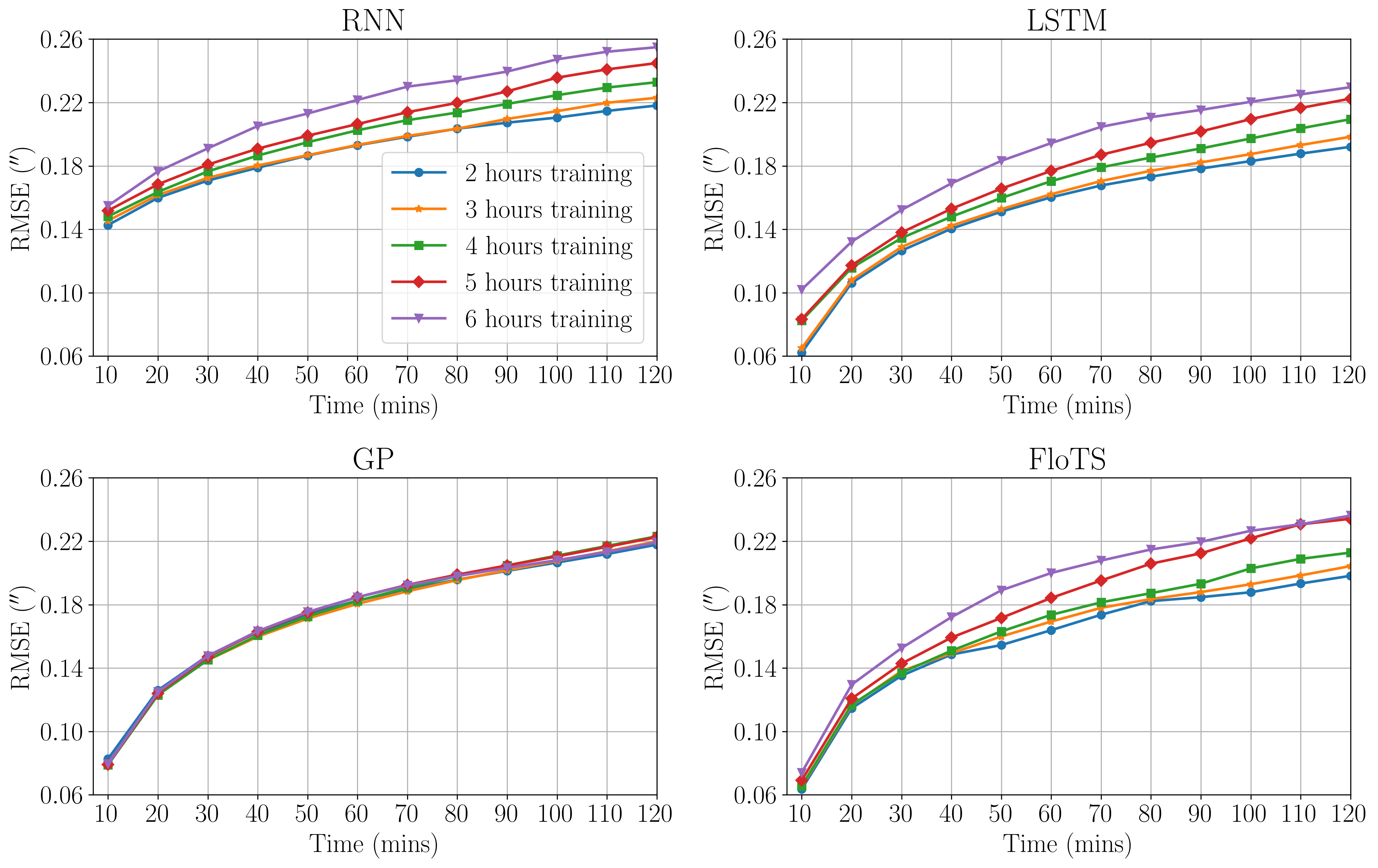}
\caption{RMSE for each training duration: 2 hours (blue), 3 hours (orange), 4 hours (green), 5 hours (red), and 6 hours (purple), evaluated over a fixed 2-hour prediction window. Results are shown for all models: RNN, LSTM, GP, and \textsc{FloTS} (FLOW).}
\label{fig:comparaison}
\end{figure*}

Figure~\ref{fig:comparaison} compares model performance by computing the RMSE between the predicted seeing values (\(\hsp\)) and the observations (\(\sp\)) across different historical data lengths (2, 3, 4, 5, and 6 hours) for a 2-hour prediction. As expected, the RMSE increases with prediction horizon across all models due to growing uncertainty and the weaker influence of initial conditions. 
The main result of this comparison is that extending the input history beyond 2 hours does not improve the predictive skill in the present univariate setting. For the neural models, longer input histories even lead to a slight degradation of performance, suggesting that older seeing measurements do not provide additional useful information for the 2-hour forecast horizon considered here. This behavior is consistent with a relatively short effective temporal correlation in the seeing time series. It may also reflect the increased difficulty of training sequence models on longer inputs, as well as the limited ability of the simple GP kernel used here to exploit multi-scale temporal dependencies.

A clear trend is observed for the RNN, LSTM, and \textsc{FloTS} models: as the historical data length decreases, the RMSE  also decreases, with the best performance achieved for the 2-hour historical data. This suggests that shorter historical data lengths limiting the complexity of the temporal relationships the model must learn, reducing overfitting, and improving its ability to generalize. Additionally, shorter input windows align more closely with the 2-hour prediction target, allowing the model to focus on the most relevant temporal patterns without being influenced by distant, less impactful data points.  That said, using longer historical windows does not degrade performance by much; the RMSE curves remain consistent across training lengths and still show a clear decrease as the input window shortens. We therefore retain the 2-hour window as the default, as it yields the lowest RMSE while keeping the input compact and well matched to the prediction horizon.

For the GP model, varying the historical data length does not significantly affect prediction performance, as shown in Figure \ref{fig:comparaison} by the overlapping curves for all input lengths. This behavior can be attributed to the nature of the Gaussian Process formulation and its kernel. Once the hyperparameters are learned, the GP relies on the kernel function to model temporal correlations between observations. In practice, this means that predictions are primarily influenced by data points that lie within the kernel’s effective correlation length.
As a result, increasing the amount of historical data beyond this correlation range does not significantly affect the predictions. The GP becomes relatively insensitive to the total length of the input window, since distant observations contribute very little to the prediction due to their low covariance.
Consequently, even when trained on different durations of data, the GP produces nearly identical performance for a fixed prediction horizon, as it effectively exploits only the most relevant, locally correlated information.

Consequently, to maintain consistency across models, we selected a 2-hour training window as the optimal choice for the 2-hour prediction horizon. This training duration was also adopted in the studies by \citet{Milli}, \citet{Turchi_2022}, and \citet{Masciadri_2}.

\subsection{Calibration of Probability Distributions}
\label{sec:calibration}

\begin{figure*}
\centering
\includegraphics[width=.8\linewidth]{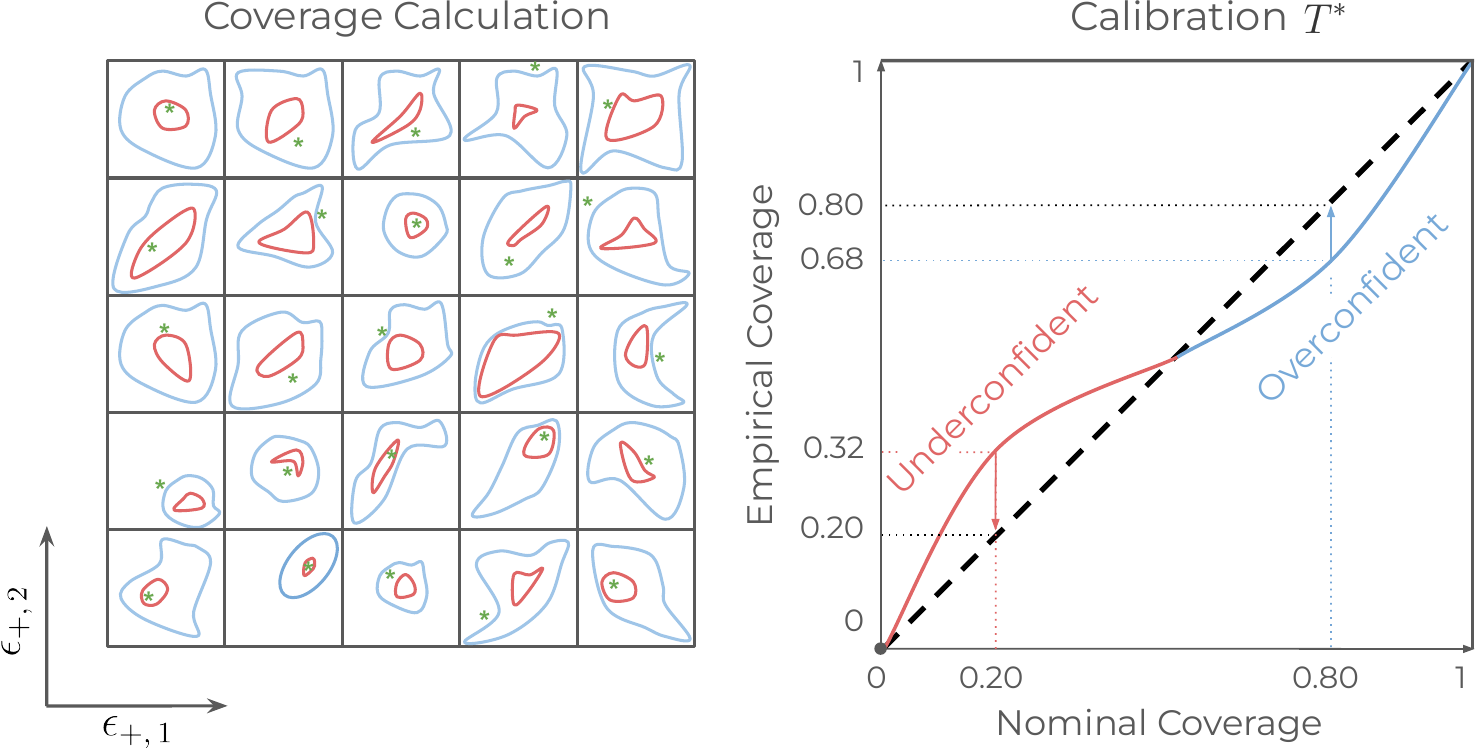}
\caption{Illustration of the calculation and calibration of the coverage for an example: predictions for 10 and 20 minutes. \textbf{Left:} examples of different observations (green star) and the corresponding predicted probability distribution shown by two contours of nominal coverage 0.2 (inner, red) and 0.80 (outer, blue). \textbf{Right:} the corresponding uncalibrated PP-plot of the empirical vs nominal coverage. The dashed black diagonal line is the ideal coverage, and the red (blue) curve represents under-(over) confident regions. The two figures together illustrate that the contours of nominal coverage 0.2 (0.80) have an empirical coverage of 0.32 (0.68), indicating under-(over) confidence. The ideal calibration temperature vector, a function of the nominal coverage, contracts (expands) the probability density of under-(over) confident regions in the parameter space.} 
\label{fig:cov_diagram}
\end{figure*}

\begin{figure}
\centering
\includegraphics[width=\linewidth]{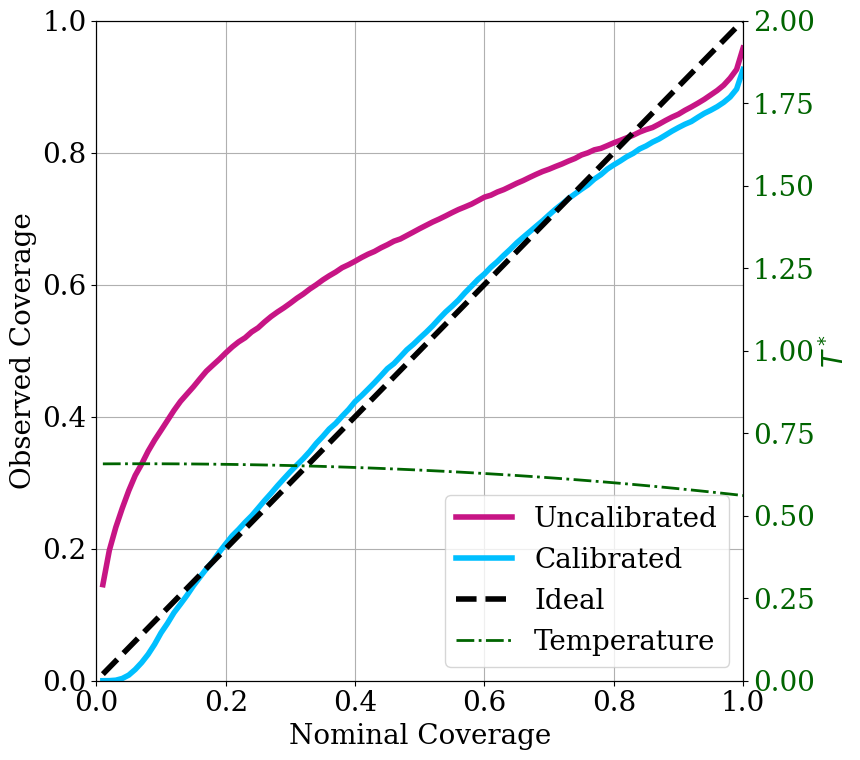}
\caption{Comparison of the PP plots from the uncalibrated (purple) and calibrated (blue) \textsc{FloTS} predictions. The 45$^\circ$ black dashed line represents the ideal coverage. The green dot-dashed line, which corresponds to the secondary y-axis, shows the optimal temperature as a function of nominal coverage. The calibration temperature is parameterized according to Equation~\ref{eq:calib_temp_polynomial}.}
\label{fig:coverage_flow}
\end{figure}

\begin{figure}
\centering
\includegraphics[width=\linewidth]{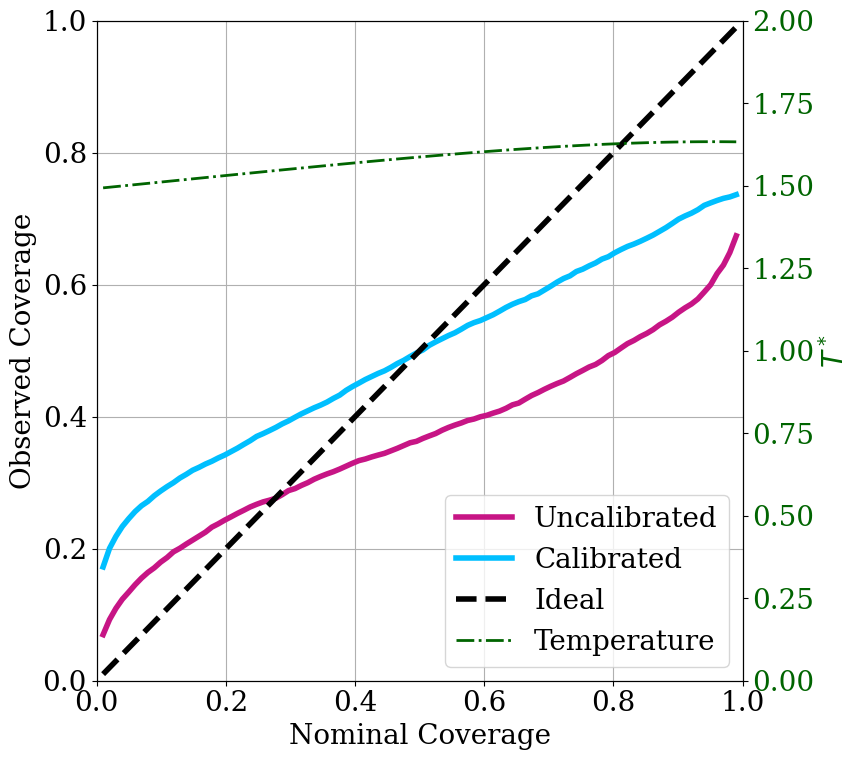}
\caption{Same as Figure~\ref{fig:coverage_flow} for GP predictions. An alternative calibration is presented in Appendix~\ref{sec:appendix_piecewiselinear}.}
\label{fig:coverage_gp}
\end{figure}

Ideally, in the limit of infinite data samples, expressivity, and computational resources, normalizing flows can accurately model probability distributions. Likewise, assuming Gaussianity, GPs can also correctly model the distribution. However, in practice, due to the finite amount of data and limited expressivity of the network/kernel, discrepancies may arise between the model approximation and the true underlying distribution. Since normalizing flows and GPs are density estimators that enable rapid sample generation and corresponding probability estimation, it is feasible to efficiently evaluate the quality of the predicted distribution and apply fine corrections through calibration.

The coverage is the probability with which a credible or confidence region contains the observed value across multiple realizations of the dataset. A parameter region with credibility $a$ is a subset of the parameter space wherein the true value lies with a probability $a$. We define the credible region using the highest posterior density (HPD), which by construction is the smallest region with this credible probability. To compute the credible regions, we compute the inverse cumulative distribution function ($F^{-1}$) from the samples, in descending order of probability density. The nominal coverage (i.e., the HPD credible region) at level $a_0$ is the set of $\sp$ values for which $p(\sp|\sm)$ is greater than or equal to the threshold $p(F^{-1}(a_0))$, such that the cumulative posterior mass does not exceed $a_0$. The empirical coverage is the proportion of times the true value falls within this credible region of the corresponding expected, nominal coverage. A well-calibrated density estimator will have matching empirical and nominal coverages.

Building on the calibration prescription of \cite{Srinivasan+25_LISA_GBs}, a generalization of the temperature scaling in \cite{Guo+2017}, we look at the mismatch between the empirical and nominal coverage to characterize the fidelity of the modeled probability distribution with that of the data samples. To illustrate the potential mismatch, we use a probability-probability (PP) plot, as illustrated in Figure~\ref{fig:cov_diagram}, which displays the empirical coverage as a function of the nominal credibility levels. A perfectly calibrated model yields a diagonal $45^\circ$ line due to the perfect agreement between expected and nominal coverage. Deviations above (below) this line reveal under(over)-confidence in the predictive distributions as illustrated in Figure~\ref{fig:cov_diagram}.

As depicted in Figure~\ref{fig:cov_diagram}, in the likely scenario of under- or over-confident empirical coverage for different values of the nominal coverage, we apply a temperature-based calibration technique shown below,
\be
\centering
p^*(\boldsymbol{\sp} \vert \boldsymbol{\sm}) = p(\boldsymbol{\sp} \vert \boldsymbol{\sm}) ^ {1/\boldsymbol{T}^*(\boldsymbol{a})},
\label{eq:calib}
\ee

where $p^*(\sp \vert \sm)$ and $p(\sp \vert \sm)$ are the calibrated and uncalibrated probability distributions respectively. $\boldsymbol{T}^*$ is the calibration temperature vector as a function of the nominal coverage $\boldsymbol{a}$ and which minimizes the mean square difference between the ideal PP plot ($45^\circ$ line) and that computed by scaling the uncalibrated probability density. For a given nominal coverage $a_0$, if the empirical coverage is under-confident ($>a_0$), the corresponding calibration temperature $\boldsymbol{T}^*(a_0)<1$ to contract the distributions. Conversely, if the empirical coverage is over-confident ($<a_0$), $\boldsymbol{T}^*(a_0)>1$ to expand the distributions. A perfectly calibrated distribution will have $\boldsymbol{T}^*(a) = 1$. 

For our current analysis, we find adequate calibration when approximating $\boldsymbol{T}^*$ as a bi-quadratic function of the nominal coverage given by
\be
\centering
\boldsymbol{T}^*(\boldsymbol{a}) = T_5\boldsymbol{a}^4 + T_4\boldsymbol{a}^3 + T_3\boldsymbol{a}^2 + T_2\boldsymbol{a} + T_1,
\label{eq:calib_temp_polynomial}
\ee
where $T_1,\,T_2,\,T_3,\,T_4,\,T_5$ are fitting coefficients that minimize the mean square difference between the ideal and uncalibrated PP plots, computed using a stochastic gradient descent algorithm.

In practice, we produce samples (i.e., predictions) from a calibrated distribution by importance-weighted resampling of the uncalibrated samples with weights given by the ratio of the calibrated and uncalibrated probability densities described in Equation~\ref{eq:calib}.

We present the uncalibrated and calibrated PP plots for \textsc{FloTS} and GP in Figures~\ref{fig:coverage_flow} and \ref{fig:coverage_gp}, respectively. Notably, for the GP, the calibrated curve remains significantly offset from the ideal 45$^\circ$ line, indicating a mismatch between the predictive distribution and the empirical distribution of the seeing.

While the Gaussian predictive assumption likely contributes to this behavior, it is not the only possible source of miscalibration. Atmospheric seeing can exhibit heteroskedasticity, non-stationarity, and regime mixing, since the training set combines observations obtained under different meteorological conditions and turbulence regimes. These effects can lead to predictive errors whose variance and distribution depend on time and atmospheric state, and may therefore not be fully captured by a stationary Gaussian GP with a simple kernel. The observed calibration mismatch should thus be interpreted as the result of a combination of factors: the Gaussian predictive form, the simplified GP configuration adopted in this work, and the intrinsic variability and regime-dependent structure of the seeing process.

To address this, we investigate a more flexible, nonparametric temperature parameterization in Appendix~\ref{sec:appendix_piecewiselinear}. Nevertheless, even with this increased flexibility, achieving alignment with the ideal calibration curve remains challenging, suggesting that post-hoc calibration alone may not fully compensate for model misspecification or for the regime-dependent structure of the data.

\section{Results}
\label{sec:results}

\begin{figure*}
\centering
\includegraphics[width=\linewidth]{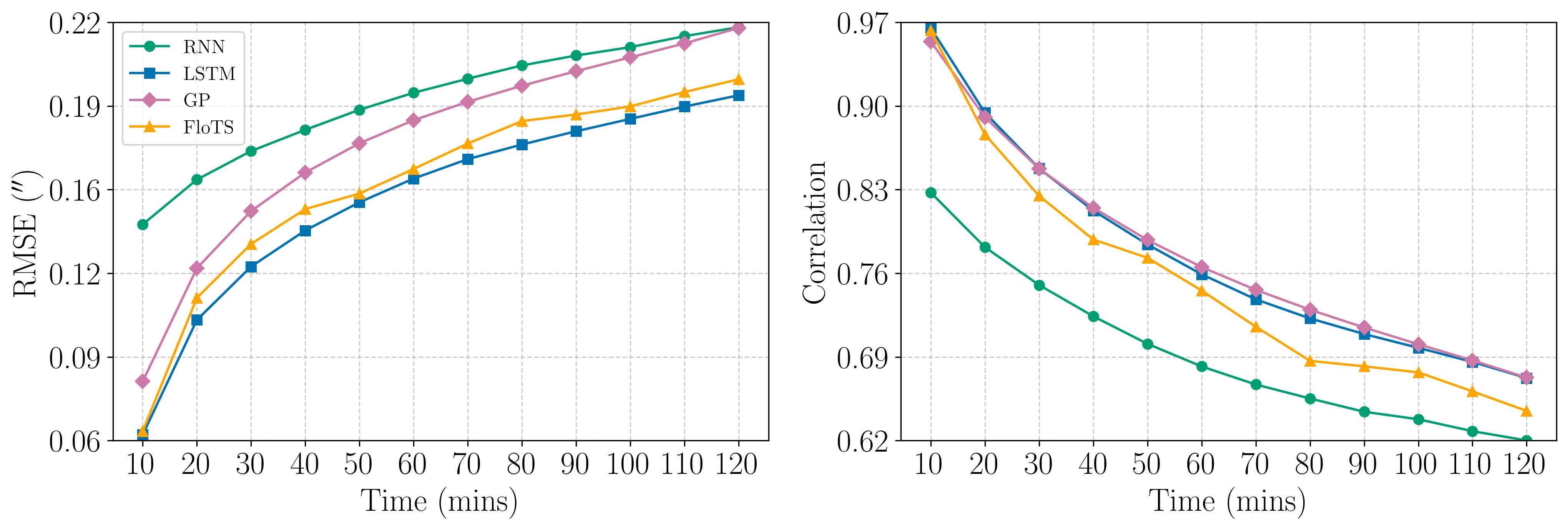}
\caption{Forecasting performance of all models trained using a 2-hour input window. \textbf{Left:} RMSE, and \textbf{Right:} Pearson correlation coefficient, both computed between the predicted and observed seeing values and plotted as a function of forecast lead time.}
\label{fig:stat_results}
\end{figure*}

Figure~\ref{fig:stat_results} compares the forecasting performance of the four models (RNN, LSTM, GP, and \textsc{FloTS}) trained with a 2-hour input duration. Overall, \textsc{FloTS} achieves lower RMSE and higher correlation across most forecast lead times, indicating more accurate and robust seeing predictions.

Across both metrics, we observe a consistent degradation of performance as the forecast lead time increases: RMSE increases, while the Pearson correlation decreases. These metrics provide complementary information on forecast quality. RMSE evaluates the typical amplitude of the prediction error in arcseconds, whereas the Pearson correlation measures how well the predicted sequence follows the temporal variations of the observations. Therefore, their joint use allows us to assess both the absolute accuracy of the forecasts and their ability to reproduce the relative evolution of the seeing over time.

In Figure \ref{fig:stat_results}, left, we show that by the 2-hour mark, RMSE values reach approximately 0.20$^{\prime\prime}$ for both LSTM and \textsc{FloTS}, and around 0.22$^{\prime\prime}$ for GP and RNN. These results indicate that LSTM and \textsc{FloTS} achieve substantially lower errors on average. On the right, the correlation coefficients show that LSTM and GP maintain the strongest correlations with the ground truth across most of the forecast window, with \textsc{FloTS} close behind, and RNN trailing in both metrics.
These results underscore the superior short-term predictive capabilities of LSTM, \textsc{FloTS}, and GP models, especially during the early stages of the forecast horizon. The LSTM benefits from its gated architecture, which allows it to selectively retain useful temporal information and mitigate the effects of noise. The \textsc{FloTS} model performs similarly well, due to its probabilistic and non-linear modeling capacity, which captures both the mean behavior and the uncertainty structure of the data. The GP slightly underperforms, suggesting that a Gaussian approximation does not fully capture the distribution of seeing values.

In summary, the combination of low RMSE and high correlation for LSTM, \textsc{FloTS}, and GP indicates both precise and trend-consistent predictions. RNN, on the other hand, lags in both respects, confirming its limited capacity to model complex temporal dependencies.
While the LSTM achieves the lowest point-prediction errors overall, and the GP offers strong short-term accuracy grounded in a well-defined statistical framework, the \textsc{FloTS} model is motivated by a different yet practically crucial criterion: the ability to deliver uncertainty estimates beyond point forecasts. 

Standard LSTM inference typically provides a single prediction, whereas both GP and \textsc{FloTS} produce probabilistic forecasts.
However, the nature of this probabilistic output differs substantially. A GP yields a predictive distribution that is Gaussian, typically summarised by a mean and variance. This is a powerful and principled choice, but it also imposes a specific distributional shape. In practice, real-world time series often exhibit departures from Gaussianity.
In contrast, the flow-based formulation underlying \textsc{FloTS} learns a flexible conditional distribution directly from data. This allows it to represent non-Gaussian predictive shapes that vary across regimes, while remaining competitive in point accuracy. As a result, \textsc{FloTS} provides a richer and potentially more realistic quantification of uncertainty, making it particularly valuable for applications where probabilities are as important as average error.

 To further illustrate the advantages of the \textsc{FloTS} model, especially its ability to characterize predictive uncertainty, we present below a case study. There, we visualize the predicted density from the \textsc{FloTS} model alongside the point forecasts of other models and the confidence intervals of the GP. This allows for a more intuitive appreciation of the \textsc{FloTS} model’s strengths in capturing both accuracy and uncertainty. For these reasons, we consider the \textsc{FloTS} model to be the most informative and robust overall.

\begin{figure*}
    \centering
    \includegraphics[width=.8\linewidth]{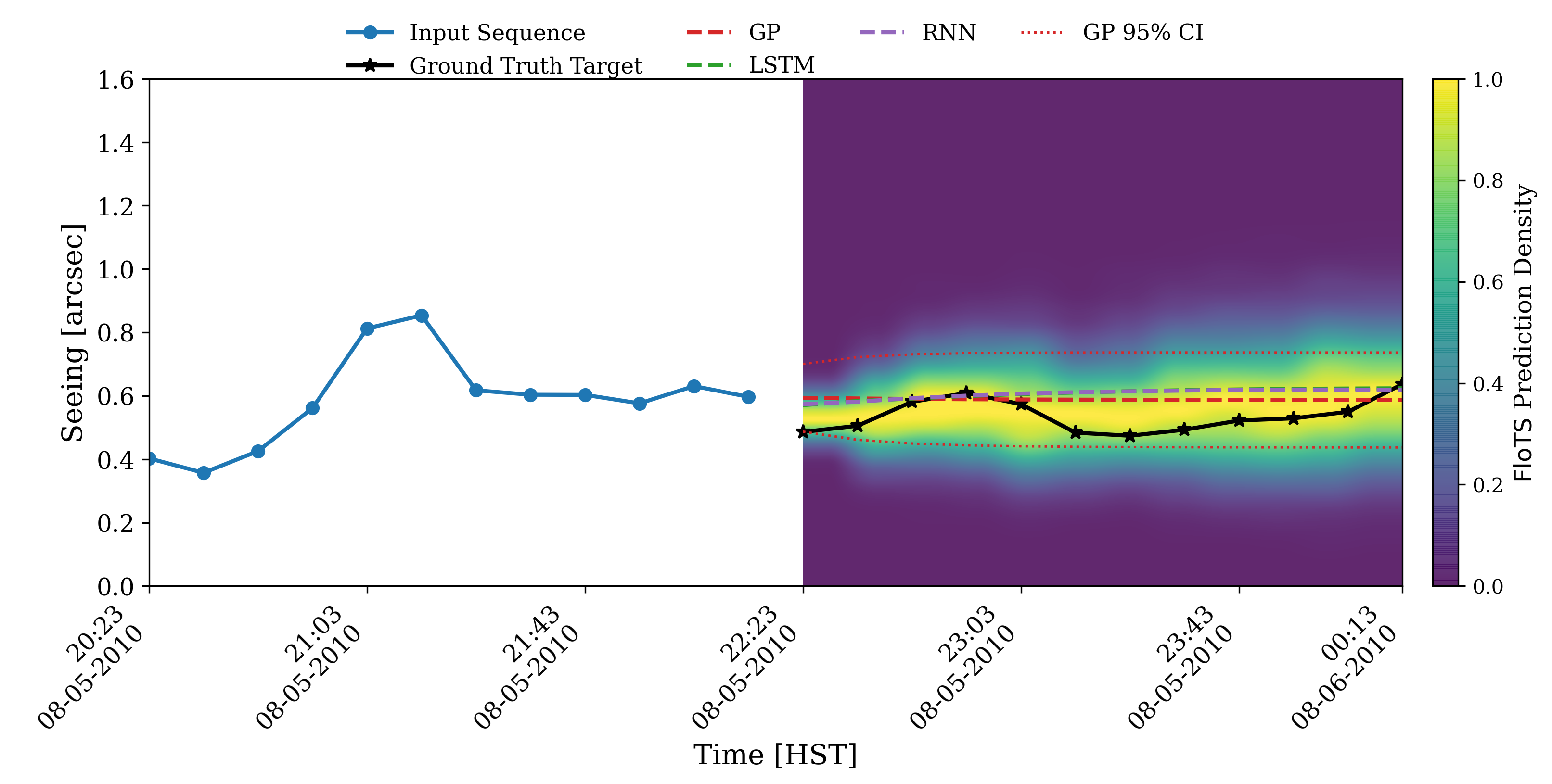}\\
    \includegraphics[width=.8\linewidth]{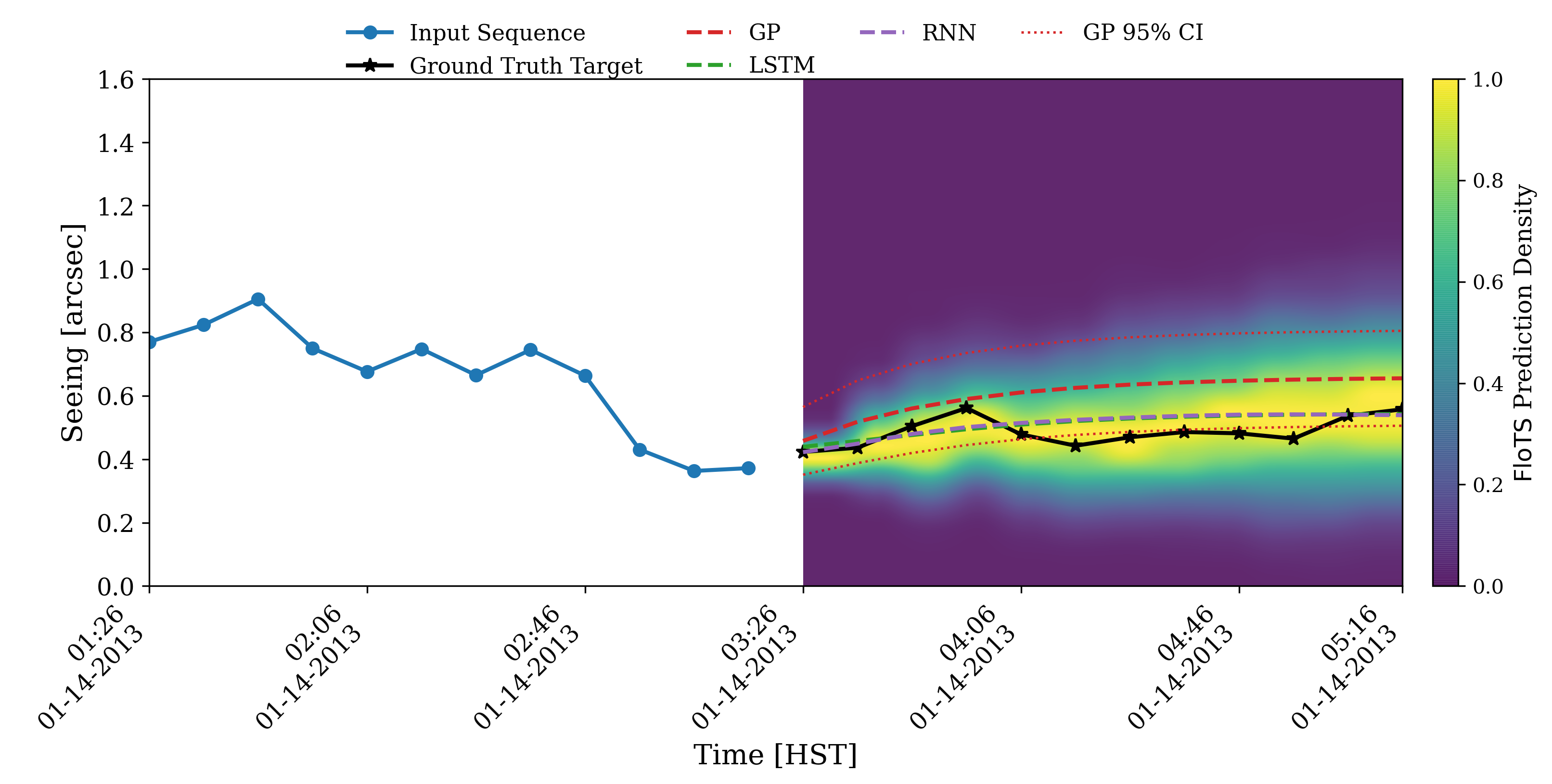}\\
    \includegraphics[width=.8\linewidth]{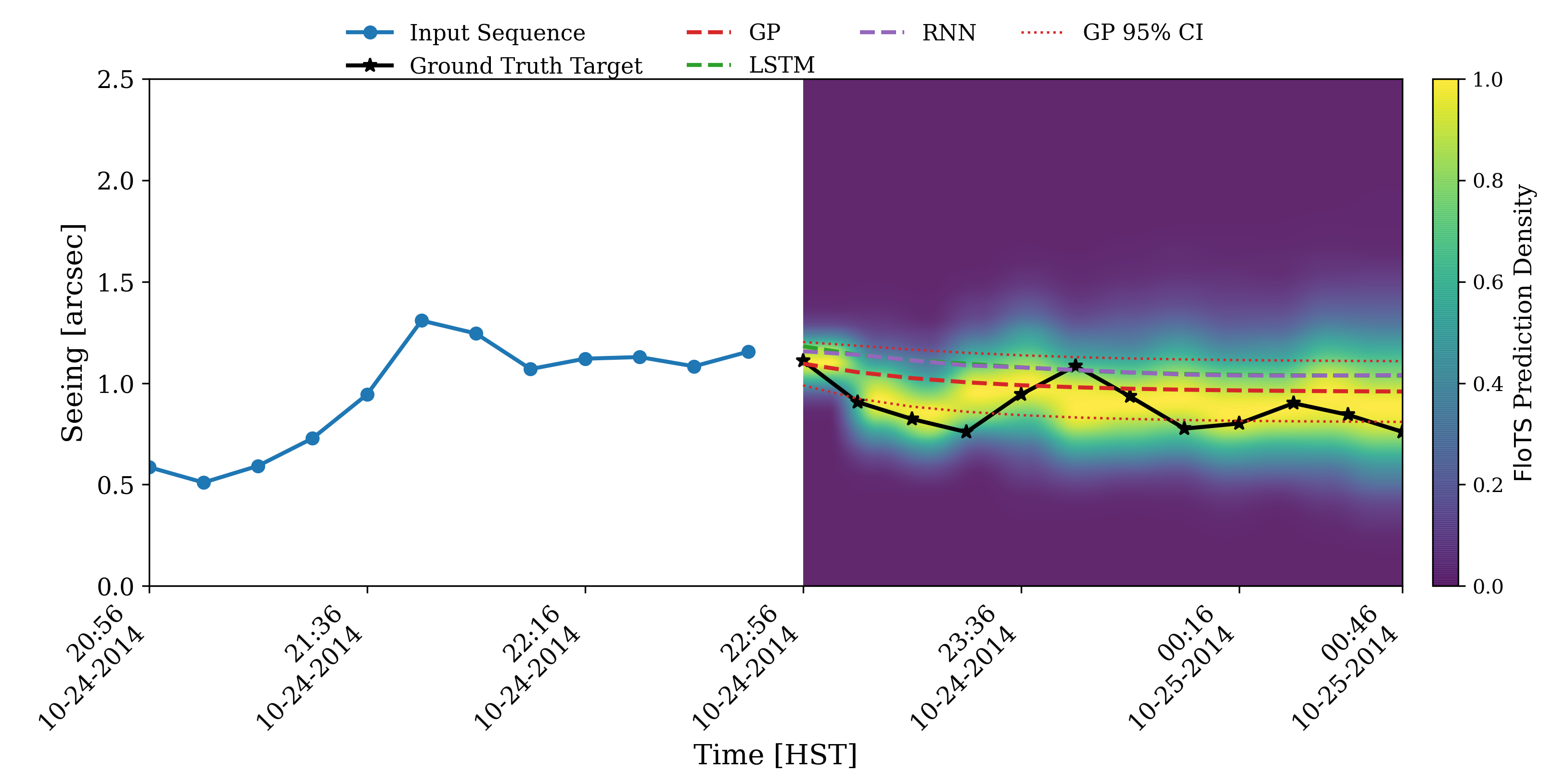}
    \caption{
    Examples of short-term seeing forecasts for three seasonally distinct and variable cases: a summer case on 05 August 2010, a winter case on 14 January 2013, and an autumn case on 24 October 2014. 
    In each panel, the left side shows the input sequence used for forecasting, while the shaded right side corresponds to the two-hour forecast horizon. The black curve indicates the observed target sequence. The red dashed curve shows the GP mean prediction. The purple and green dashed curves show the RNN and LSTM predictions, respectively. The heatmap represents the probability density predicted by \textsc{FloTS}, estimated from posterior samples and normalized at each forecast timestep for visualization.
    }
    \label{fig:cs}
\end{figure*}
\subsection*{Case Study}
\label{sec:cs}
To complement the quantitative evaluation, we present three case studies to qualitatively examine the behavior of the different forecasting models on representative examples. The examples were selected from different seasons in order to cover distinct atmospheric regimes: a summer case in August, a winter case in January, and an autumn case in October. These cases provide a challenging and operationally relevant evaluation, as they include strong temporal variability, non-monotonic behavior, and changes in the seeing regime over the forecast horizon, allowing for an intuitive comparison of the model outputs, particularly in terms of the uncertainty captured by the probabilistic \textsc{FloTS} predictions and by the GP.

Figure~\ref{fig:cs} presents the three selected examples. The first case, from August 2010, illustrates a summer night with significant short-term variability. The input sequence exhibits a marked increase in seeing, followed by a partial stabilization before the prediction window. During the forecast horizon, the observed seeing evolves non-monotonically, with a local increase followed by a decrease and then a renewed rise. In this case, the GP prediction remains relatively smooth, while \textsc{FloTS} assigns a structured predictive density around the observed evolution. The ground truth remains within regions of non-negligible to high predicted density, indicating that the flow-based model captures the uncertainty associated with this variable evolution.

The second case, from January 2013, corresponds to a winter example with a clear transition in the input sequence. The seeing decreases substantially before the forecast window and then remains variable during the target period. This case is representative of a less stationary regime, where the forecast cannot be described as a simple continuation of the previous trend. During the forecast horizon, the observed seeing first increases toward a local maximum of approximately 0.6$^{\prime\prime}$, then decreases before rising again toward the end of the prediction window. The \textsc{FloTS} density follows this evolving region of the target trajectory and broadens over the forecast horizon, while the deterministic RNN and LSTM forecasts provide smoother point estimates. The GP mean also remains smooth and captures the average tendency, but its Gaussian confidence interval cannot fully represent the more structured shape of the flow-based predictive distribution.

The third case, from October 2014, represents an autumn example with higher seeing values and stronger temporal fluctuations. 
The target sequence exhibit non-monotonic evolution with several local variations. This case is particularly useful for emperically evaluating the models under a more active regime. The GP prediction tends to revert smoothly, while \textsc{FloTS} produces a broader and more flexible predictive density, reflecting the larger uncertainty associated with the future seeing evolution. The observed target remains largely within the high-density region of the \textsc{FloTS} prediction, while the density also reflects the ambiguity induced by the variable input sequence.

Overall, these examples show that the different models behave consistently with their underlying assumptions. The RNN and LSTM provide deterministic point forecasts that generally follow the broad temporal tendency, but they tend to be smoother and less responsive to short-term variations. The GP provides an interpretable probabilistic forecast through its Gaussian predictive mean and confidence interval, but its uncertainty remains constrained by the Gaussian assumption and by the smoothness imposed by the learned covariance structure. In contrast, \textsc{FloTS} provides a full predictive density that can represent broader, asymmetric, and locally structured uncertainty. This flexibility is particularly useful in the variable regimes shown here, where the future seeing evolution is not simply a smooth extrapolation of the input sequence. 

It is important to highlight that these case studies also illustrate the scope of the univariate forecasting framework adopted in this work. Since all models are conditioned only on past seeing values, their predictions rely on the temporal information already present in the recent seeing sequence. This approach can be effective when the recent evolution remains informative about the near future, but it may become limited during regime transitions or rapidly evolving turbulence events, where external atmospheric drivers are not fully encoded in the seeing history alone. The variable cases shown here therefore provide a useful baseline for assessing how much information can be extracted from past seeing before extending the input space to include physically relevant meteorological predictors.

\begin{figure*}
    \centering
    \makebox[\textwidth]{%
    \includegraphics[width=1\textwidth]{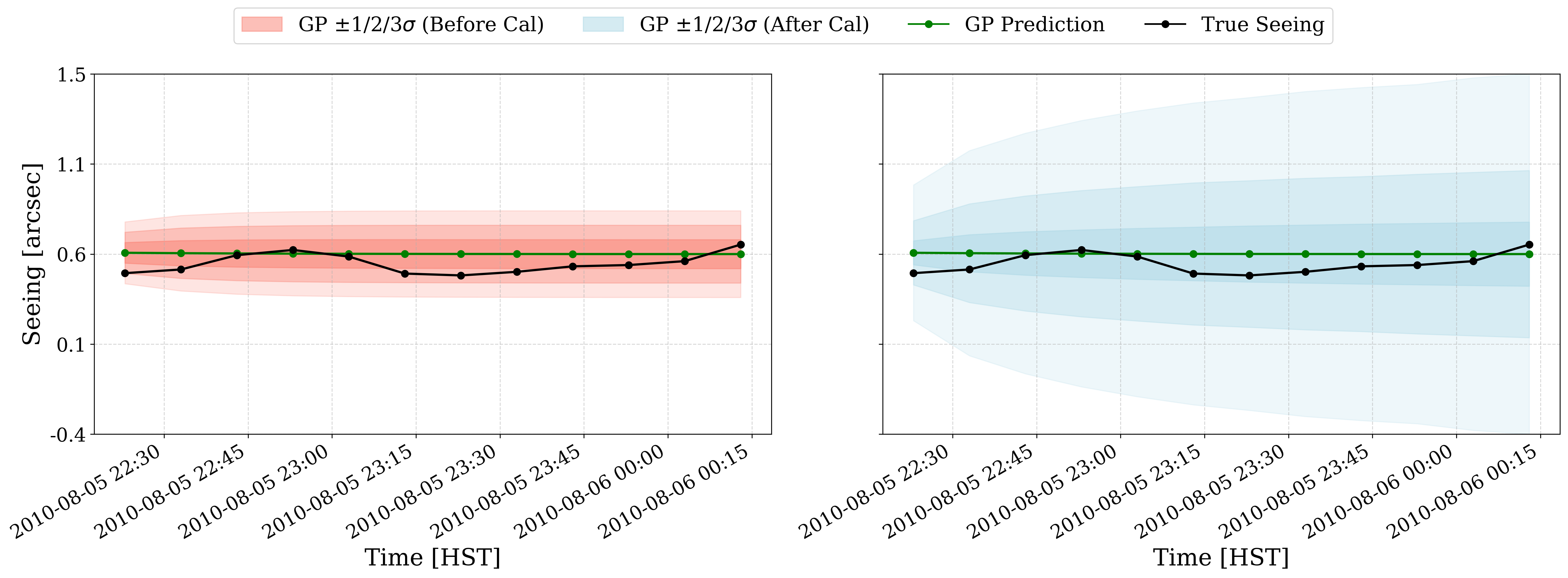}
    }
    \caption{
    Comparison of the GP predictive uncertainty before and after calibration for the August case study.
    \textbf{Left:} GP predictive mean and uncertainty bounds over the two-hour forecast horizon before calibration. The shaded red regions indicate the $\pm1\sigma$, $\pm2\sigma$, and $\pm3\sigma$ confidence intervals around the GP prediction.
    \textbf{Right:} Same prediction after calibration, with blue shaded regions denoting the calibrated uncertainty bands.
    In both panels, the green line represents the GP predictive mean, and the black markers indicate the observed seeing measurements. Calibration leaves the GP mean unchanged but increases the predictive dispersion, especially at longer lead times.
    }
    \label{fig:gp_calib}
\end{figure*}

\begin{figure*}
    \centering
    \includegraphics[width=.8\linewidth]{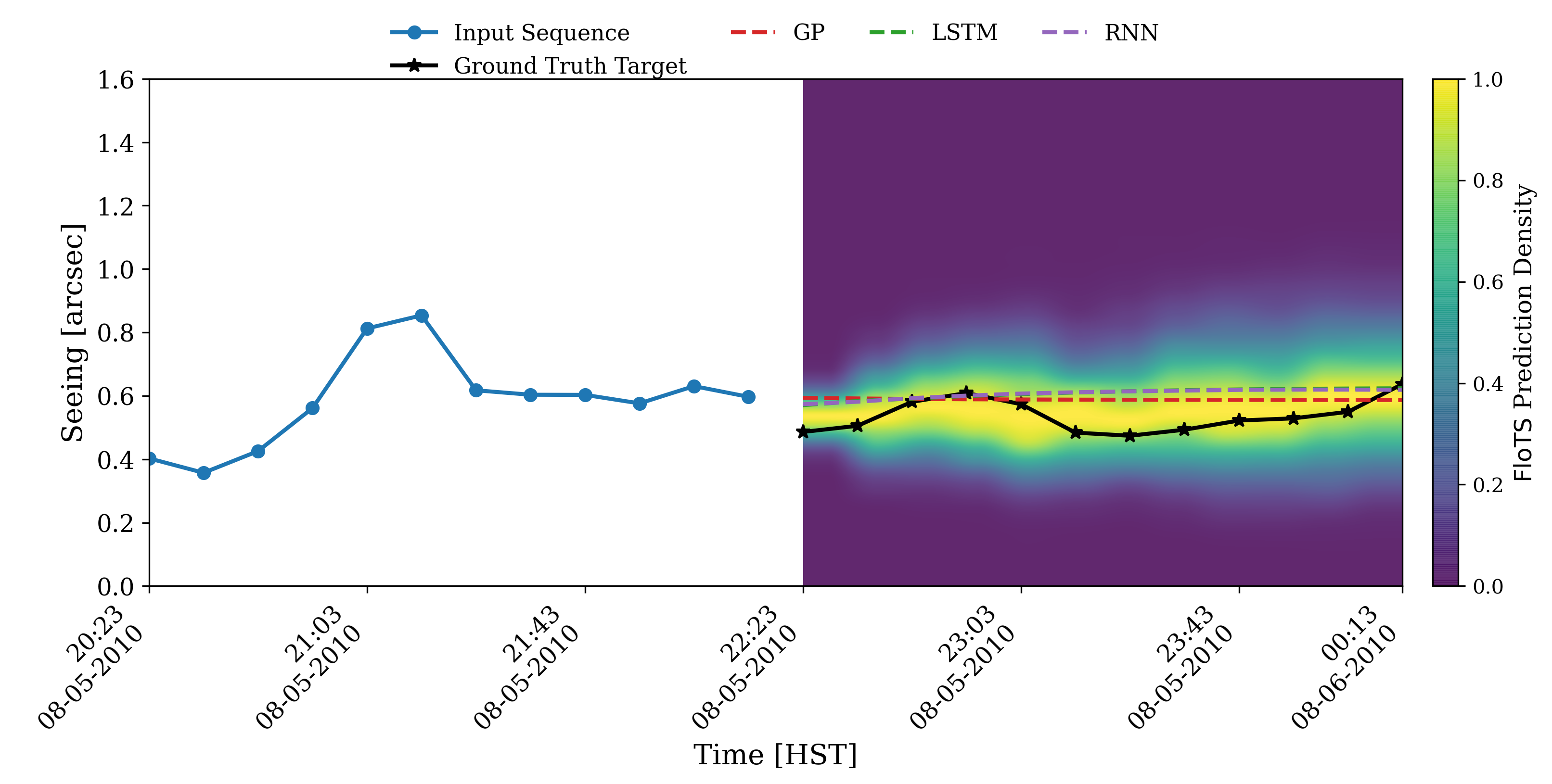}\\
    \includegraphics[width=.8\linewidth]{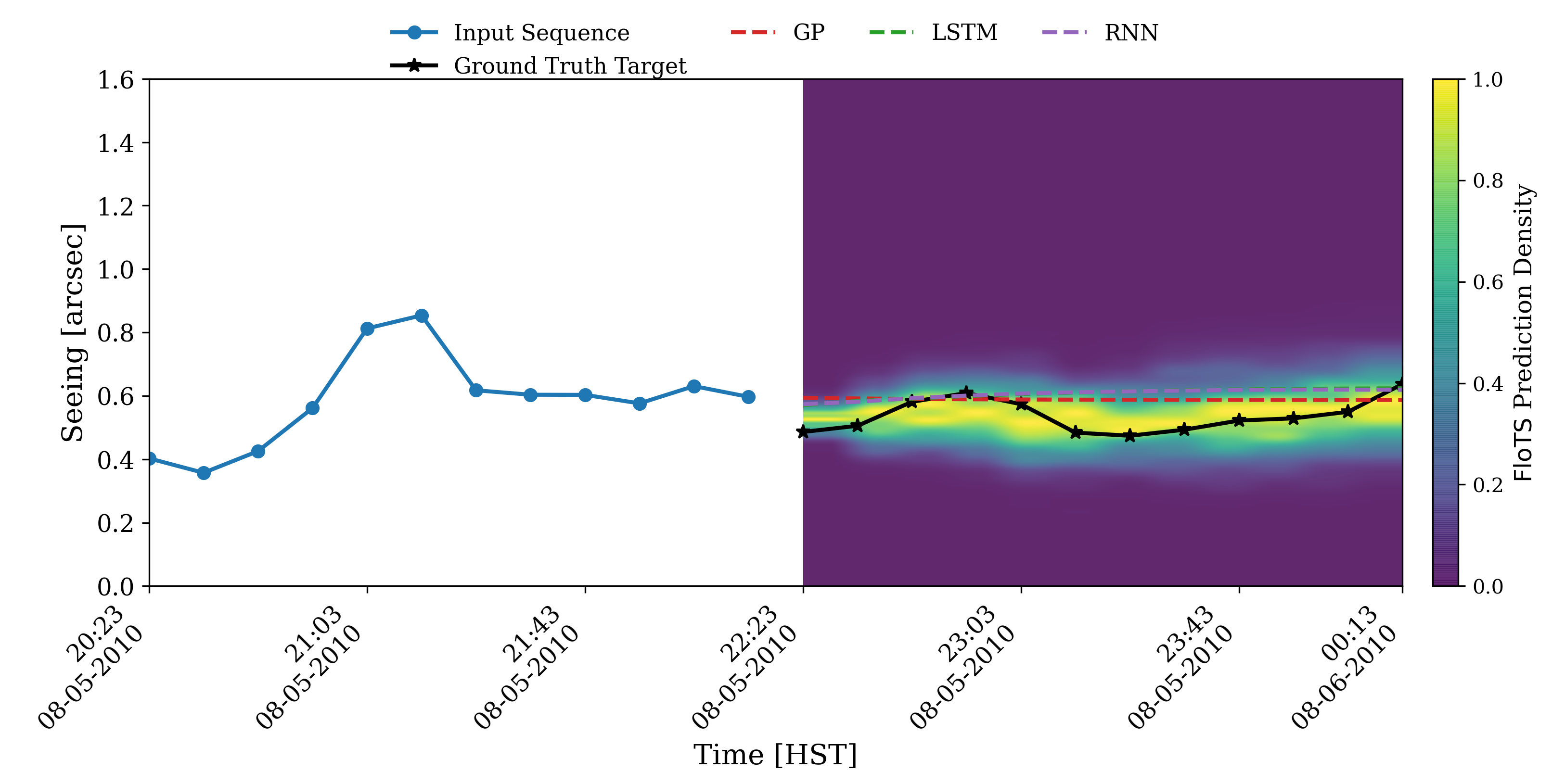}
    \caption{
    Comparison of the \textsc{FloTS} seeing prediction before and after calibration for the August case study.
    The top panel shows the uncalibrated prediction, while the bottom panel shows the calibrated prediction. In each panel, the left side shows the input sequence used for forecasting, and the right side displays the two-hour forecast horizon. The black curve indicates the observed target sequence, while the red, purple, and green dashed curves show the GP, RNN, and LSTM predictions, respectively. The heatmap represents the probability density predicted by \textsc{FloTS}. For visualization, the density is normalized independently at each forecast timestep. Calibration redistributes probability mass within the predictive density, modifying both the dispersion and the local non-Gaussian structure of the forecast distribution.
    }
    \label{fig:flow_calib}
\end{figure*}

To further highlight the importance of calibration for probabilistic forecasting, we compare the GP and \textsc{FloTS} predictions before and after calibration for the August case study 
(Figures~\ref{fig:gp_calib} and~\ref{fig:flow_calib}). For the GP, calibration leaves the predictive mean unchanged by construction and primarily modifies the dispersion of the predictive intervals. As shown in Figure~\ref{fig:gp_calib}, the calibrated uncertainty bands are wider than the uncalibrated ones, particularly at longer lead times, yielding more conservative uncertainty estimates. This behavior is consistent with the coverage analysis presented in Figure~\ref{fig:coverage_gp}, which showed that the uncalibrated GP tends to be over-confident.

Figure~\ref{fig:flow_calib} shows the corresponding uncalibrated and calibrated predictions for \textsc{FloTS}. Unlike the GP, where calibration mainly rescales the predictive variance, calibration of \textsc{FloTS} can redistribute probability mass within the full predictive density. As a result, it can modify not only the overall dispersion but also the local non-Gaussian structure of the forecast distribution. This is particularly relevant in variable regimes, where the predictive uncertainty may be asymmetric, broadened, or structured around several plausible future evolutions.

Together, these case studies demonstrate the advantages of \textsc{FloTS} probabilistic modeling for capturing both the trajectory and the uncertainty of future seeing values under seasonally diverse and active conditions. While the RNN and LSTM models provide deterministic point forecasts, and the GP provides a useful Gaussian uncertainty estimate, \textsc{FloTS} offers a richer probabilistic representation of the forecast horizon. These examples therefore complement the global quantitative metrics by illustrating the behavior of the models in regimes that are representative of non-stationary and operationally relevant seeing variability.

\section{Conclusions and Discussion} 
\label{sec:conclusion}

The objective of this study was to develop an accessible and effective method for short-term seeing prediction, with practical applications in astronomical observation and optical communication. We aim to forecast seeing conditions up to 2 hours ahead, using MKWC historical data. To this end, we evaluated four models: a statistical model, GP, and three deep learning approaches, RNN, LSTM, and a normalizing flow-based model. A key distinguishing feature of our work compared to previous studies is the integration of probabilistic models, specifically GP and \textsc{FloTS}, that not only provide deterministic forecasts but also quantify predictive uncertainty. This capability is particularly valuable for decision-making processes where understanding the confidence level of predictions is essential, for example in telescope scheduling.

In this first study, we deliberately restricted the input space to past seeing values only. This choice provides a clean and controlled framework for comparing different forecasting models under identical input conditions. It also allows us to isolate the predictive information contained in the seeing time series itself before introducing additional physical predictors. The goal of this univariate formulation is therefore not to build a complete physically informed forecasting system at this stage, but rather to establish a baseline assessment of short-term seeing predictability and to identify a suitable modeling framework for representing both the temporal evolution of seeing and its associated uncertainty.

Our experiments revealed that a 2-hour training window was optimal for all models, suggesting that, within the univariate setup considered here, the most useful predictive information is concentrated in the recent seeing history.
With this optimal context length, all four models produced reasonable forecasts, exhibiting the expected trend: increasing RMSE with forecasting horizon, while correlation with ground truth progressively decreases.

This result indicates that recent seeing measurements contain exploitable short-term information over the two-hour forecast horizon considered here. However, the univariate setup also limits the physical interpretation of the predictability measured in this work. In particular, part of the predictive skill may reflect persistence in the seeing time series, especially under quasi-stationary atmospheric conditions. Therefore, the present results should be interpreted as a benchmark of the information content of past seeing alone, rather than as a complete demonstration of predictability beyond persistence or beyond the external atmospheric drivers of optical turbulence.

Among the models, LSTM achieved the lowest errors overall, slightly outperforming the others in terms of point prediction accuracy. However, when considering both accuracy and uncertainty quantification, the \textsc{FloTS} model offers the most favorable trade-off between precision and informativeness. As shown in the error and correlation metrics, its performance is very close to that of LSTM. Importantly, the case studies demonstrate that the \textsc{FloTS} model's predictive density not only captures the general trends of the ground truth but also tracks local variations, such as short-term increases and decreases in seeing values, with higher fidelity. This capacity to model both global behavior and fine-grained fluctuations enhances the realism of the predictions, making them more consistent with the observed data.

The comparatively lower performance of the GP model should therefore be interpreted in light of the specific GP baseline adopted in this study, rather than as a fundamental limitation of Gaussian Process methods. The use of a simple stationary exponential kernel, together with hyperparameter estimation on a restricted training subset for computational efficiency, may limit the ability of the GP to represent multi-scale or non-stationary seeing dynamics. More expressive GP formulations, including multi-scale or non-stationary kernels and scalable sparse approximations, remain an important direction for future work.

Such probabilistic forecasting makes the \textsc{FloTS} model particularly well-suited for operational scenarios that demand not only accurate forecasts but also robust non-Gaussian uncertainty quantification to inform decision-making and risk assessment. The GP model follows closely, providing useful uncertainty estimates through a more constrained and less data-adaptive structure. An important observation is that the predictions from the LSTM and RNN consistently fall within the high-probability regions of the GP and \textsc{FloTS} predicted densities, reinforcing the reliability of their uncertainty estimates. Furthermore, the 95\% confidence intervals of the GP model consistently enclose the peaks where the \textsc{FloTS} model maintains moderate to high confidence, emphasizing the complementary nature of these two probabilistic approaches. These results also highlighted the importance of calibration for probabilistic forecasts, especially for \textsc{FloTS}, whose predictive distribution is inherently non-Gaussian. This leads to more reliable and trustworthy predictions.

The scope of the univariate approach should therefore be understood in relation to the atmospheric regime being forecast. A model based only on past seeing values is expected to perform best when the atmosphere is relatively quasi-stationary, such that the recent temporal evolution remains informative about the near future. In contrast, during rapidly evolving conditions, intermittent turbulence events, or transitions between atmospheric regimes, past seeing alone may not contain sufficient information to anticipate future changes. In those situations, additional physically relevant predictors are likely required.

In future work, we aim to investigate the incorporation of additional meteorological features, such as temperature, wind speed and direction, humidity, and atmospheric pressure, which are known to influence optical turbulence. This extension is not only a natural next step for improving forecast performance, but also a necessary step to determine whether the remaining performance limitations arise from model capacity or from the intrinsic information content of the univariate seeing input. In particular, wind shear, thermal stratification, humidity variations, pressure changes, and large-scale flow conditions may provide information that is not contained in the past seeing sequence alone, especially during non-stationary or regime-transition events. While these variables may provide useful contextual information, their actual contribution to model performance remains to be evaluated. Feature engineering and selection will be necessary to determine whether these additional inputs enhance the predictive accuracy or simply add noise to the system \citep{dong2018feature}. 

Beyond the recurrent baselines considered in this thesis, self-attention-based Transformer \citep{wolf2020transformers} architectures could be explored as an alternative sequence model. In contrast to RNNs/LSTMs, which compress past information into a fixed-size hidden state, Transformers form predictions by weighting past time steps through attention. This mechanism may be advantageous for seeing time series where the relevant temporal dependencies are not strictly local and may occur at multiple, potentially irregular, time scales. More generally, other generative architectures such as conditional variational autoencoders (VAEs) could also be considered for probabilistic forecasting, as they learn a conditional latent representation of the future given past observations \citep{vae}.

Furthermore, future research could explore ensemble strategies that leverage the strengths of different models, for instance, combining the LSTM’s deterministic forecasting capabilities with the \textsc{FloTS} model’s probabilistic uncertainty quantification. These extensions would move toward a more comprehensive and operationally useful forecasting framework, capable of delivering both accurate predictions and confidence estimates, while also accounting for physical drivers of seeing variability.

\section*{Acknowledgements}
The authors thank the Maunakea Weather Center (MKWC), particularly Dr. Steven Businger, Dr. Tiziana Cherubini, and Ryan Lyman, for providing essential atmospheric data that supported this study. We also acknowledge the Seeing Monitor Group for their efforts in maintaining and providing access to seeing measurements. Additionally, we extend our appreciation to the Canada-France-Hawaii Telescope (CFHT) staff for their operational support in maintaining the instruments and database that contributed to the seeing observations.
RS acknowledges support from the European Union’s Horizon ERC Synergy Grant “Making Sense of the Unexpected in the Gravitational-Wave Sky” (Grant No. GWSky-101167314). Computational work has been partly made possible through SISSA-CINECA and CINECA-INFN agreements providing access to resources on LEONARDO at CINECA.MB. We gratefully acknowledge funding from the Centre National d’Études Spatiales (CNES) and the Université Côte d’Azur (UCA). This work was also supported by the French government through the UCAJEDI “Investments in the Future” program, managed by the National Research Agency (ANR) under grant ANR-15-IDEX-01.

\newpage
\appendix

\section{Piecewise linear calibration}
\label{sec:appendix_piecewiselinear}
\begin{figure}
\centering
\includegraphics[width=.5\linewidth]{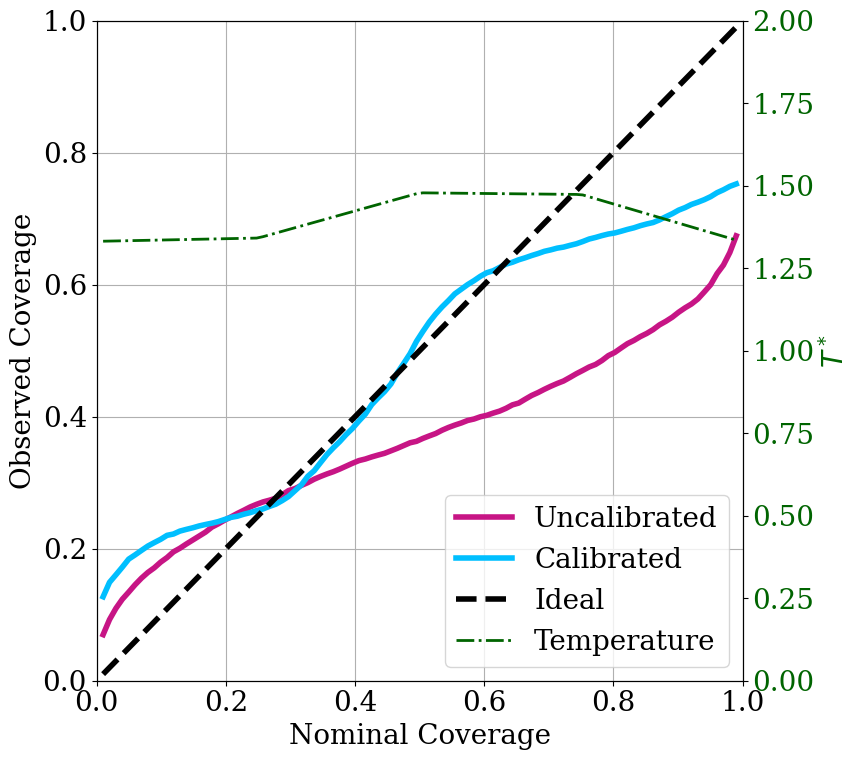}
\caption{Comparison of the PP plots from the uncalibrated (purple) and calibrated (blue) GP predictions. The 45$^\circ$ black dashed line represents the ideal coverage. The green dotted-dashed line, corresponding to the secondary y-axis, shows the optimal temperature as a function of nominal coverage. The calibration temperature is fit according to Equation~\ref{eq:calib_temp_piecewiselinear}.}
\label{fig:coverage_piecewiselinear_GP}
\end{figure}

In addition to the smoothly varying polynomial parameterization of the temperature as a function of nominal coverage shown in Equation~\ref{eq:calib_temp_polynomial}, we also consider a non-smooth alternative based on a piecewise-linear representation with the same number of free parameters. This formulation enables flexible localized adjustments of the calibration curve. Specifically, the function $\boldsymbol{T}^*(\boldsymbol{a})$ is defined as a piecewise-linear interpolation between five learnable anchor temperatures ${T_k}$ at fixed uniformly spaced nodes,
\be
\boldsymbol{T}^*(\boldsymbol{a}) =
\begin{cases}
T_1 + \dfrac{T_2 - T_1}{0.25}(a), & a \in [0, 0.25), \\[6pt]
T_2 + \dfrac{T_3 - T_2}{0.25}(a - 0.25), & a \in [0.25, 0.5), \\[6pt]
T_3 + \dfrac{T_4 - T_3}{0.25}(a - 0.5), & a \in [0.5, 0.75), \\[6pt]
T_4 + \dfrac{T_5 - T_4}{0.25}(a - 0.75), & a \in [0.75, 1.]. \\[6pt]
\end{cases}
\label{eq:calib_temp_piecewiselinear}
\ee
In Figure~\ref{fig:coverage_piecewiselinear_GP}, we present the piecewise-linear calibration of the GP. Compared to Figure~\ref{fig:coverage_gp}, this approach yields a modest improvement in median coverage. However, as with the polynomial calibration, discrepancies persist at both low and high nominal coverage levels, and exact calibration is not achieved. This further underscores the limitation of the Gaussian assumption in GP models for describing the underlying seeing distribution, and explains the better calibration performance of the flow model shown in Figure~\ref{fig:coverage_flow}.

\section{Posterior predictive distributions from GP}
\label{sec:appendix}
\begin{figure*}
    \centering
    \includegraphics[width=.8\linewidth]{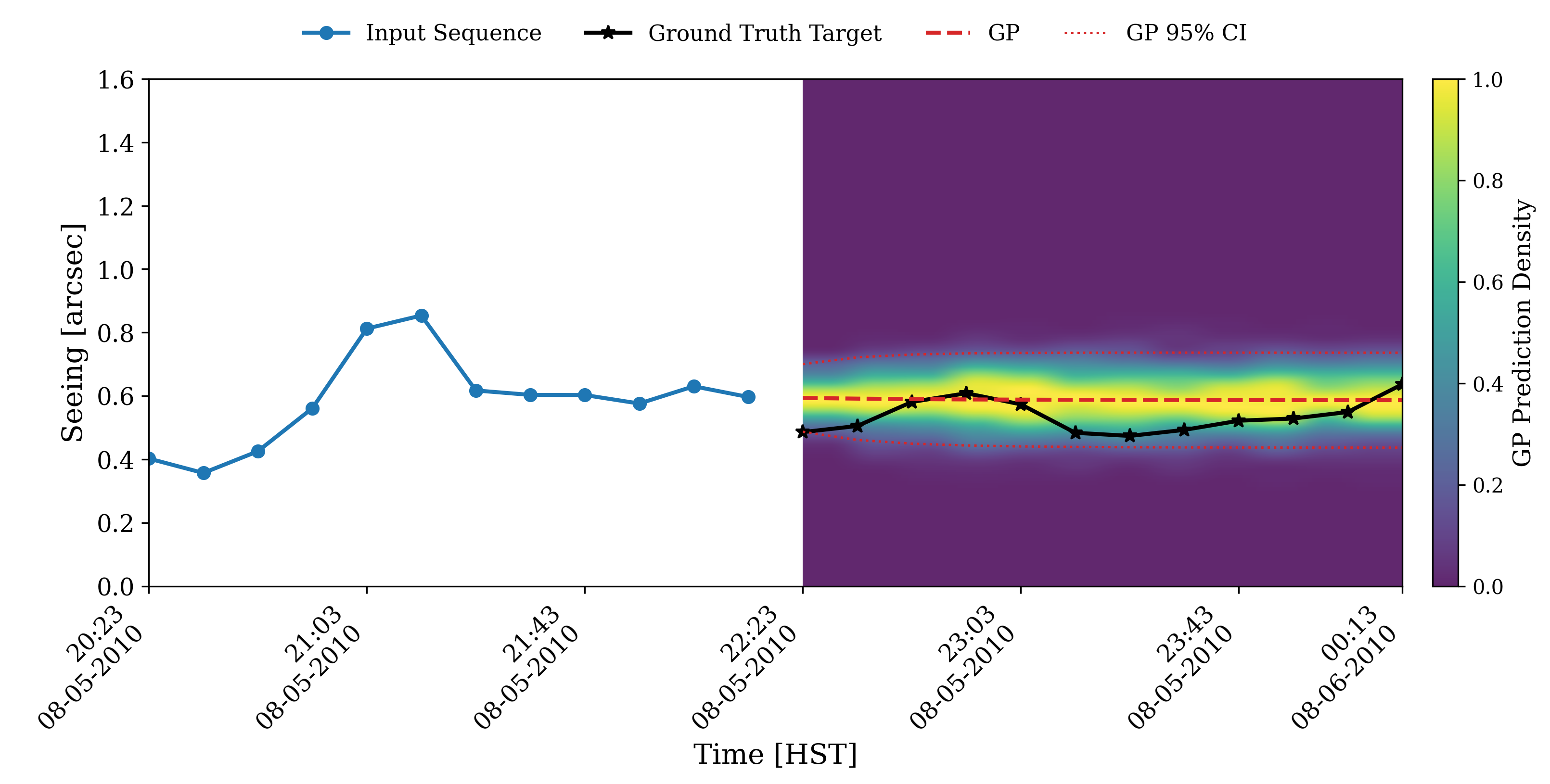}\\
    \includegraphics[width=.8\linewidth]{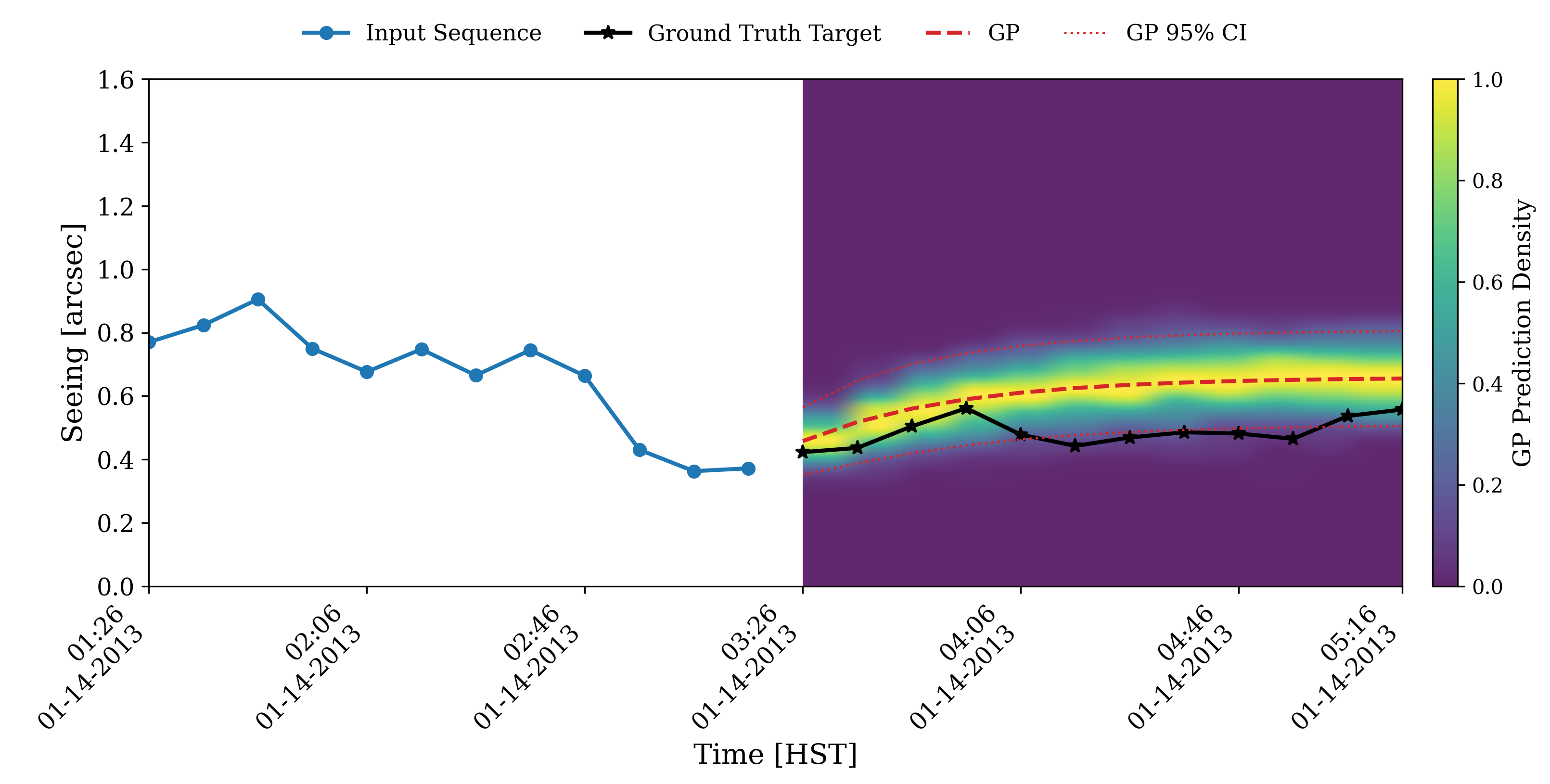}\\
    \includegraphics[width=.8\linewidth]{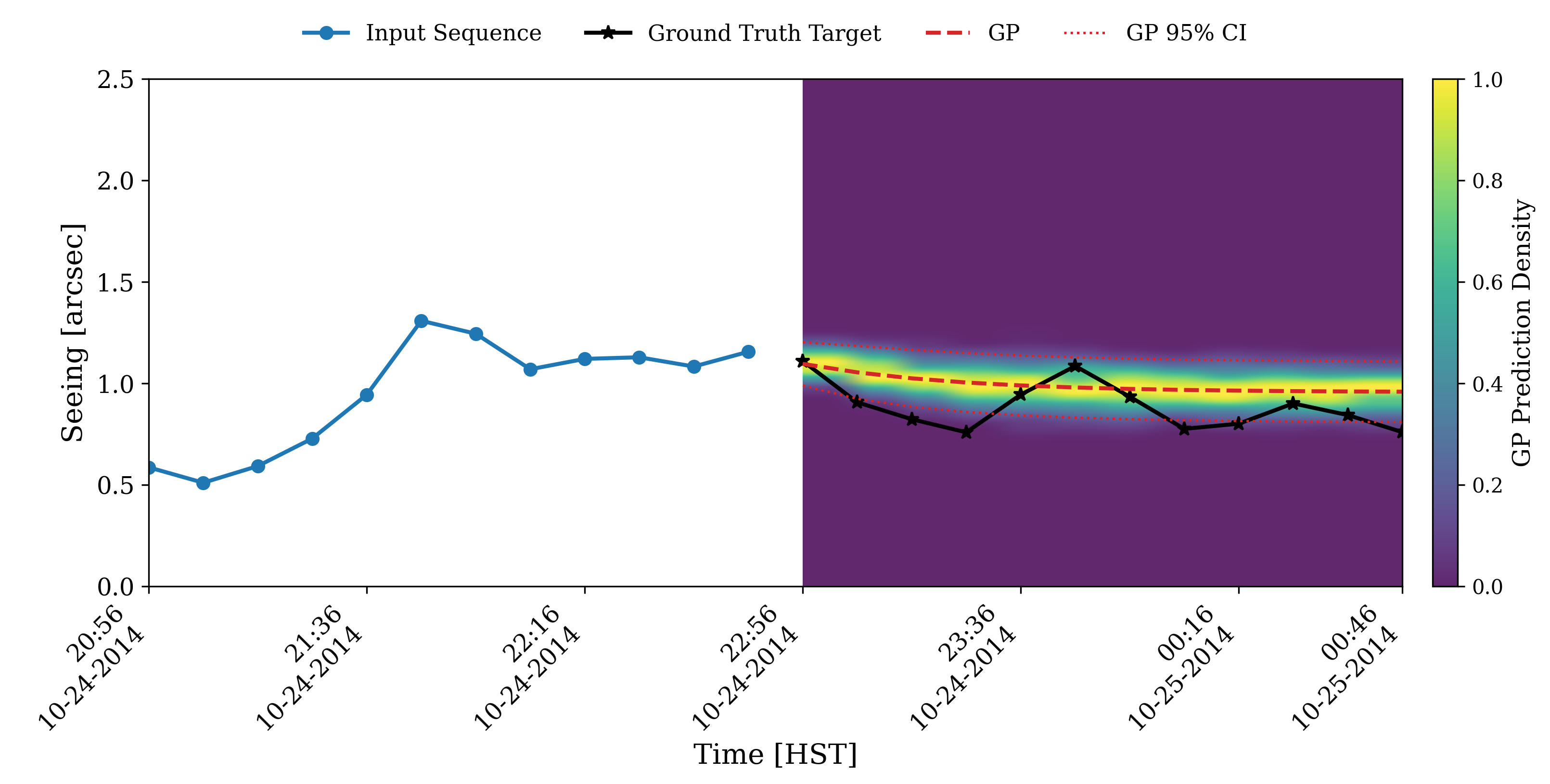}
    \caption{
    GP posterior predictive distributions for the same three case studies shown in Figure~\ref{fig:cs}: 
    a summer case on 05 August 2010, a winter case on 14 January 2013, and an autumn case on 24 October 2014.
    The posterior mean is shown as a red dashed line, while the 95\% confidence interval is indicated by small red dotted lines. The input sequence is shown in blue, and the ground-truth target values are shown in black. The heatmap represents the probability density inferred from 5000 GP posterior samples.
    }
    \label{fig:gp_samples}
\end{figure*}

Figure~\ref{fig:gp_samples} presents the uncalibrated posterior predictive distributions from the GP model for the same case studies discussed in Section~\ref{sec:cs}. These plots illustrate the uncertainty representation produced by the GP and provide a direct comparison with the \textsc{FloTS} predictive densities shown in Figure~\ref{fig:cs}. 

As expected from the GP formulation, the posterior predictive density is smooth and largely determined by the predictive mean and covariance. The sampled trajectories are therefore distributed around the posterior mean according to the Gaussian uncertainty structure of the model. This provides an interpretable representation of uncertainty, but it also limits the range of shapes that the predictive distribution can express.

In comparison with \textsc{FloTS}, the GP density maps exhibit less local structure and cannot easily represent asymmetric or more complex non-Gaussian uncertainty patterns. This difference is particularly visible in the more variable case studies, where the future seeing evolution is not simply a smooth continuation of the input sequence. These posterior predictive plots therefore highlight the complementary behavior of the GP and \textsc{FloTS}: the GP provides a smooth Gaussian uncertainty estimate, while \textsc{FloTS} can represent more flexible, data-driven predictive densities.

\label{lastpage}

\bibliography{ref}{}
\bibliographystyle{aasjournal}

\end{document}